\documentclass[twocolumn]{jpsj2} 
\def\d{{\rm d}}

\def\kf{k_{\rm F}}

\def\ggs{\buildrel\textstyle > \over {\hbox{\raise0.2ex\hbox{$\sim$}}}}
\def\lls{\buildrel\textstyle < \over {\hbox{\raise0.2ex\hbox{$\sim$}}}}

\def\im{{\rm i}}

\def\et{{\it et al.}}
\def\delx{\partial_x}

\def\delx{\partial_x}

\def\delx{\partial_x}
 
\def\jo #1#2#3#4{#1 {\bf #2} (#3) #4}   
\def\PR{Phys.\ Rev.}
\def\PRB{Phys.\ Rev.\ B}
\def\PRL{Phys.\ Rev.\ Lett.}

\def\SSC{Solid State Commun.}

\def\JCP{J.\ Chem.\ Phys.}
\def\JPC{J.\ Phys.\ C}

\def\JPIF{J.\ Phys.\ I\ France}
\def\JDP{J.\ de\ Phys.}
\def\JPCON{J.\ Phys.\ Condens. Matter}

\def\JPSJ{J.\ Phys.\ Soc.\ Jpn.}

\def\RMP{Rev.\ Mod.\ Phys.}

\def\SOV{Sov.\ Phys.\ JETP}

\def\EPL{Europhys.\ Lett}

\def\SM{Synth.\ Met.}

\def\BCSJ{Bull.\ Chem.\ Soc.\ Jpn.}
\def\JPCS{J.\ Phys.\ Chem.\ Solids}
\def\CR{Chem.\ Rev.} 
\def\CL{Chem.\ Lett.} 
\def\JACS{J.\ Am.\ Chem.\ Soc.} 
\def\CPL{Chem.\ Phys.\ Lett.} 

\title
{Theoretical Aspects of Charge Ordering in Molecular Conductors}

\author{Hitoshi \textsc{Seo}$^1$\thanks{E-mail address: seo@post.kek.jp},  
Jaime \textsc{Merino}$^2$, 
Hideo \textsc{Yoshioka}$^3$, 
and Masao \textsc{Ogata}$^4$} 

\inst{$^1$
Non-Equilibrium Dynamics Project, ERATO-JST, 
c/o KEK, Tsukuba 305-0801, Japan\\ 
$^2$Departamento de F\'isica Te\'orica de la Materia 
Condensada, Universidad Aut\'onoma de Madrid, Madrid 28049, Spain\\
$^3$Department of Physics, Nara Women's University, Nara 630-8506, Japan\\
$^4$Department of Physics, Faculty of Science, University of Tokyo, Tokyo 113-0033, Japan
}

\recdate{\today}
\abst{Theoretical studies on charge ordering phenomena 
in quarter-filled molecular (organic) conductors are reviewed. 
Extended Hubbard models
including not only the on-site but also the inter-site Coulomb repulsion
are constructed in a straightforward way from the crystal structures, 
which serve for individual study on each material 
as well as for their systematic understandings. 
In general the inter-site Coulomb interaction stabilizes 
Wigner crystal-type charge ordered states, where the charge localizes 
in an arranged manner avoiding each other, and can drive the system insulating. 
The variety in the lattice structures, represented by anisotropic networks 
in not only the electron hopping but also in the inter-site Coulomb repulsion,  
brings about diverse problems in low-dimensional strongly correlated systems. 
Competitions and/or co-existences 
between the charge ordered state and other states are discussed, 
such as metal, superconductor, and the dimer-type Mott insulating state which is another typical 
insulating state in molecular conductors. 
Interplay with magnetism, 
e.g., antiferromagnetic state and spin gapped state for example due to the spin-Peierls transition, 
is considered as well. 
Distinct situations are pointed out: 
influences of the coupling to the lattice degree of freedom 
and effects of geometrical frustration which exists in many molecular crystals.  
Some related topics, such as charge order in transition metal oxides 
and its role in new molecular conductors, are briefly remarked. 
}
\kword{molecular conductors, charge ordering, strongly correlated electron system, 
low-dimensional systems, metal-insulator transition, 
superconductivity, magnetism, electron-lattice coupling, geometrical frustration}

\begin{document}
\maketitle

\section{Introduction}\label{sec_Intro}

\vspace{3pt}

The charge ordering (CO) phenomenon 
is actively studied in the research field of charge transfer type molecular conductors~\cite{CR}, 
since it plays a key role in their physical properties. 
Following its clear observation in DI-DCNQI$_2$Ag~\cite{Hiraki98PRL}, 
metal-insulator transitions in many of these materials 
are now understood as due to CO. 
It has been found not only in newly synthesized compounds but also 
in systems which have been well known for years, 
where, however, its existence has been veiled until recently. 
Typical examples are 
two representative families of this field, 
the Bechgaard salts TM$_2X$~\cite{Chow00PRL} 
(TM = TMTSF or TMTTF) 
and ET$_2X$~\cite{Miyagawa00PRB} 
(ET = BEDT-TTF) where $X$ takes different atoms or molecules, 
both being studied for more than 20 years. 
The CO transition was revealed recently and 
now became one of the most important issues in these systems. 

These materials are members of 
the so-called 2:1 salts expressed as $A_2B$, 
to which interest of this field has been conducted. 
Numerous compounds with such a 2:1 composition have been synthesized 
and found to exhibit a rich variety of properties~\cite{Farges94Book,Ishiguro98Book,Bernier99Book}. 
Many of them show electron conduction at room temperature, 
where the carriers are due to a charge transfer from cations $B^+$ or anions $B^-$, 
resulting in an average valence of $-1/2$ or $+1/2$ for $A$ molecules, respectively. 
The $B$ ion has closed shell in most cases 
then the valence band near the Fermi energy is composed of the frontier orbital, 
LUMO or HOMO, of the $A$ molecule, 
which is quarter-filled as a whole in terms of electrons or holes. 
The variety in their properties has been revealed to be originated from 
the diversity of anisotropic lattices 
resulting in different non-interacting band structures, 
together with strong correlation effects experienced by 
electrons among this HOMO/LUMO band determining the low energy properties~\cite{Seo04CR}. 
CO is a typical consequence of such strong correlation, 
namely, large electron-electron Coulomb repulsion compared to the kinegic energy, 
especially due to the long-range nature of this Coulomb force. 
In fact, it is now ubiqitously found in $A_2B$ compounds 
as well as in other strongly correlated electron systems 
such as transition metal oxides.

In this article, 
we review theoretical aspects of CO mainly aiming at following points: 
In what kind of situation are they formed? In what situation do they melt? 
How does the spin degree of freedom act? 
Are there any superconducting (SC) state near the CO phase? 

Such theoretical works on CO have been done from early days, 
motivated by experiments. 
For example, the metal-insulator transition in 
a classical transition metal oxide Fe$_3$O$_4$, the magnetite,   
was proposed to be due to CO by Verwey~\cite{Verwey41Physica}, 
although its existence is still controversial to date~\cite{Fe3O4}. 
Another trigger was an early molecular conductor TTF-TCNQ, 
where an incommensurate $4k_{\rm F}$ charge-density-wave (CDW) 
is observed, but only in a diffusive manner therefore long ranged order 
is not achieved~\cite{Devreese79Book,Kagoshima89Book}.
Here we refer to the term CO as the phenomenon due to strong Coulomb interaction, 
sometimes called as ``Wigner crystal on lattice"~\cite{Hubbard78PRB}. 
This should be distinguished with other transitions 
resulting in periodic modulations of the charge density, 
such as the $2k_{\rm F}$ CDW (the Peierls-Fr\"{o}lich state) driven by 
the nesting of the Fermi surface together with the electron-lattice coupling~\cite{Kagoshima89Book}, 
which is essentially a phenomenon at weak correlation. 

The observations of CO transition in $A_2B$ molecular systems 
have stimulated many theoretical studies adhered to these compounds. 
Experiments have fortunately appeared 
around when researchers in this field started to realize that 
effects of electron correlation 
could be modeled in a straightforward way and 
that rather simple models would successfully describe 
their physical properties~\cite{RKato99Book}. 
That is, 
each constituent $A$ molecule is represented by a ``site" 
and only the frontier orbital is considered. 
The non-interacting band structures near the Fermi level
are well reproduced by 
the extended H{\"u}ckel tight-binding scheme~\cite{Hoffman63JCP,TMori84BCSJ}. 
Then, the Coulomb interaction between electrons in this orbital 
is taken into account~\cite{Kino96JPSJ,Fukuyama00Physica}. 
We call such model as the extended Hubbard model (EHM), 
written as follows: 
\begin{align}
 {\cal H}_{\rm EHM} &=-\sum_{\langle ij \rangle\sigma}\left(t_{ij}c^{\dagger}_{i\sigma}
    c_{j\sigma}+h.c.\right) \nonumber \\
 & \hspace{1.5cm} +\sum_{i} U n_{i\uparrow}n_{i\downarrow}
    +\sum_{\langle ij \rangle}V_{ij} n_in_j. 
\label{eqn:extHub}
\end{align}
Here, $\langle ij \rangle$ denotes pairs of the lattice sites (i.e., molecules) $i$ and $j$, 
$\sigma$ is the spin index which takes $\uparrow$ and $\downarrow$, 
$n_{i\sigma}$ and  $c^{\dagger}_{i\sigma}$ ($c_{i\sigma}$) denote 
the number operator and the creation (annihilation) operator for the 
electron of spin $\sigma$ at the $i$th site, respectively, 
and $n_i=n_{i\uparrow}+n_{i\downarrow}$.  
The transfer integrals, $t_{ij}$, reflect the anisotropy resulting 
from the particular spatial extent of the frontier orbital, 
calculated, e.g., by the extended H{\"u}ckel method 
or from tight-binding fitting of first principle calculations. 
The Coulomb interactions of not only on-site $U$ 
but also inter-site $V_{ij}$ are considered, the latter being 
crucial for the CO as we will see later explicitly. 
In the $[A_2]^{-}B^+$ ($[A_2]^{+}B^-$) systems 
the non-interacting band as a whole is quarter-filled in terms of electrons (holes), 
namely, there exists one electron (hole) per two sites on average.  

Because of this clear way of constructing microscopic models from crystal structures,  
results of theoretical works could be checked back in the experiments, 
and such interplay has greatly accelerated the research. 
Throughout these we have learned that, 
although the basic picture of CO is rather classical and essentially known from the early days, 
the physics therein is rich and diverse. 
The main reason for such diversity is the variety in the geometry of 
lattice structures of the materials where it is realized. 
More specifically, relative positions between molecules 
not only reflect directly on the anisotropy in $t_{ij}$ controlling the band structure, 
but also affect drastically the nature of the CO state itself through $V_{ij}$.  
This is in contrast with  
the case of the Mott insulating state in half-filled systems where the driving force 
is the on-site term $U$, which is a character of the atom/molecule itself; 
in the case of the dimer-type Mott insulator in $A_2B$ systems (see later), 
it is the ``on-dimer" Coulomb repulsion, $U_{\rm dimer}$~\cite{Kino96JPSJ,Kino95JPSJ_2,Miyagawa04CR}.

The research of CO is still continuing and rather rapidly growing. 
New phenomena are uncovered, even in the above mentioned compounds,
and now under extensive investigations. 
For example 
a pressure-temperature phase diagram of 
DI-DCNQI$_2$Ag has been explored where an anomalous 
temperature dependence of metallic resistivity $\rho \propto T^3$ is found just beyond 
the border of the CO phase~\cite{Itou04PRL}. 
A peculiar interplay between CO and magnetic properties 
found in TMTTF$_2X$ compounds under pressure~\cite{Yu04PRB} requires 
a reformulation of the generic phase diagram of TM$_2X$~\cite{Jerome82AdvPhys}. 
Problems of CO system on anisotropic triangular lattice structures 
characteristic of ET$_2X$ compounds 
have many new aspects~\cite{Miyagawa00PRB,HMori98PRB}. 
We cannot offer comprehensive explanation for each of these cases here 
as many are still not yet resolved theoretically, 
but we hope that this review, by explaining the present status, 
would provide a base for tackling such new physics and lead the readers toward challenges. 

The organization of this paper is as follows. 
Since most of the $A_2B$ compounds have low-dimensional structures, 
we will devide this review into one-dimensional (1D) and two-dimensional (2D) problems, 
while in reality there exist finite interchain/interlayer interactions; 
they are quasi-1D and quasi-2D systems. 
Theoretical results devoted for the quasi-1D compounds are described in 
$\S$~\ref{sec_q1d}. 
This starts with studies on purely 1D electronic models and then additional effects, 
such as interchain interaction and coupling to the lattice degree of freedom, 
are considered.
The 1D case can be studied in more controlled ways, analytically 
and numerically, than in the study of the quasi-2D compounds, 
which we discuss in $\S$~\ref{sec_q2d}. 
There, theoretical works on 2D electronic models are still in progress 
and influence of additional effects, e.g., coupling to the lattice, is not fully understood yet. 
Studies aimed at SC states near the CO phase in 2D models, 
motivated by its observations, are reviewed as well. 
A problem of CO systems under geometrical frustration 
which is expected to be relevant to many molecular conductors will be pointed out in 
$\S$~\ref{sec_q1d} and \ref{sec_q2d}. 
In $\S$~\ref{sec_related} related topics will be mentioned, 
such as analogous CO states observed in transition metal oxides. 
Possible roles of CO in other molecular systems will be added as well, 
as perspectives. A summary is given in $\S$~\ref{sec_sum}. 

Some of the experimental studies on CO are mentioned in this review 
but many references including important ones are left out; 
we refer to other reviews from experimental standpoints~\cite{CR,TakahashiThisVol} 
which would be complementary to this article. 
One can find many review articles 
on the properties of molecular compounds in general~\cite{CR,Farges94Book,Ishiguro98Book,Bernier99Book}, 
especially we refer to refs.~\ref{Seo04CR} and \ref{Fukuyama00Physica} 
for papers from theoretical but more systematic point of views
including the CO systems as well.

\vspace{3pt}

\section{Quasi-One-dimensional Systems}\label{sec_q1d}

\vspace{3pt}

Early theoretical works on CO in 1D models have been 
performed motivated by the observation of 4$k_{\rm F}$ CDW in 
TTF-TCNQ 
as mentioned in $\S$~\ref{sec_Intro}, 
where the importance of the long-range Coulomb interaction was 
emphasized~\cite{Hubbard78PRB,Ovchinnikov73Sov,Lee77PRB,Kondo78JPSJ,Mazumdar83PRL,Hirsch83PRL}.  
However for the quasi-1D $A_2B$ systems, before CO was found,  
analyses were mainly concentrated on the Hubbard-type models 
only considering the on-site Coulomb energy $U$, 
but some did discuss the relevance of the inter-site $V_{ij}$~\cite{Mila93EPL,Penc94PRB,Mila95PRB}. 
For example, Mila~\cite{Mila95PRB} estimated that 
the nearest-neighbor Coulomb repulsion 
is appreciable in TM$_2X$, such as more than one third of $U$, 
by comparing calculations on the dimerized version of 1D EHM  (see eq. (\ref{eqn:extHub1D})) 
with optical measurements~\cite{Jacobsen86JPC}. 
After CO was found, 
many further works devoted to the 1D EHM and its variants have been carried out, 
which we will review in this section.

\subsection{Charge ordering in quasi-one-dimensional systems}\label{subsec_COq1d}

\vspace{3pt}

When the first observation of CO in DI-DCNQI$_2$Ag was made by $^{13}$C-NMR~\cite{Hiraki98PRL}, 
in an independent theoretical work based on mean-field (MF) approximation, 
Seo and Fukuyama~\cite{Seo97JPSJ} proposed that 
CO due to the nearest-neighbor Coulomb repulsion might exist in TMTTF$_2X$. 
This was based on a comparison between self-consistent solutions at zero temperature 
obtained by the standard MF treatment applied to the appropriate dimerized 1D EHM 
and the spin structure in the antiferromagnetic phase
suggested by a $^1$H-NMR measurement~\cite{Nakamura95SM}. 
Soon after, it was found that members of TMTTF$_2X$ indeed show CO, 
directly seen in a $^{13}$C-NMR measurement~\cite{Chow00PRL}. 
Moreover a divergence of the dielectric constant at the CO transition temperature is observed, 
suggesting a ferroelectric state~\cite{Monceau01PRL}. 
This is a consequence of CO, which we will mention later in this section. 

These systems are members of 
DCNQI$_2X$ ($X$: monovalent metal cation $X^+$, e.g., Ag and Li)~\cite{Hunig04CR,Hiraki96PRB}  
and TM$_2X$ ($X$: monovalent anion $X^-$, e.g., PF$_6$, AsF$_6$, SCN, and Br)~\cite{Jerome82AdvPhys}, 
respectively,  
both having quasi-1D structures. 
DCNQI stands for the $R_1R_2$-DCNQI molecule 
where $R_1,R_2$ are sustituents such as CH$_3$, Br, I, etc.
Here, quasi-1D ``structure" implies twofold meanings: 
in their crystal structures the DCNQI/TM molecules assemble in a stacking manner, 
and, in their electronic structures the transfer integrals in the interchain direction, 
$t_\textrm{inter}$, 
are one order of magnitude smaller than those in the intrachain direction, $t_\textrm{intra}$. 

\begin{figure}
\centerline{\includegraphics[width=7.5truecm]{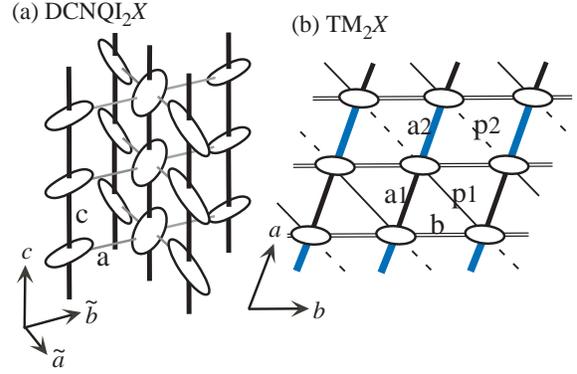}}
\vspace*{-1em}
\caption{(Color online) Schematic representation of the molecular networks
in (a) DCNQI$_2X$ and (b) TM$_2X$. The crystallographic axes 
are shown; in (a) we take $\tilde{a}=a+b$, $\tilde{b}=a-b$ 
where $a$ and $b$ are the axes in the tetragonal cell. 
The interchain interactions are three-dimensional in (a), 
whereas it is two-dimensional in (b) 
since interlayer couplings are one order of magnitude 
smaller due to the anion layers. 
The indices (for TM$_2X$, taken from ref.~\ref{TMori84BCSJ})
are not only for $t_{ij}$ but also for different values of $V_{ij}$, 
although the degree of anisotropy can be different (see text). 
}
\vspace*{-1.5em}
\label{fig_DCandTM}
\end{figure}

The networks of DCNQI/TM molecules are schematically shown in Fig.~\ref{fig_DCandTM}.
In DCNQI$_2X$ the interchain couplings are three-dimensional (3D) and uniform, 
while TM$_2X$ they are quasi-2D and the TM molecules are connected in a rather complicated way 
(the $\beta$-type structure; see $\S$~\ref{subsec_COq2d}). 
For DCNQI$_2X$ systems, $|t_\textrm{intra}| \simeq 0.15 \sim 0.25$ eV, 
$|t_\textrm{inter}| \simeq 0.01 \sim 0.03$ eV~\cite{Kashimura95SSC,Miyazaki96PRB}, 
while for TM$_2X$ systems: in TMTSF compounds $|t_\textrm{intra}|\simeq 0.2 \sim 0.4$ eV, 
$|t_\textrm{inter}| \simeq 0.01 \sim 0.05$ eV 
and in TMTTF compounds $|t_\textrm{intra}|\simeq 0.1 \sim 0.25$ eV, 
$|t_\textrm{inter}| \simeq 0.01 \sim 0.03$ eV~\cite{TMori82CL,Grant83JDP,Ducasse86JPC,TMori84BCSJ}.
These values depend on the actual salts and different methods of calculation also provide 
varied estimations. 

We note that the distances between molecules does not simply correspond to    
the degree of anisotropy in $t_{ij}$, 
since the anisotropic shape of the frontier orbitals makes the dependence of $t_{ij}$ to the 
relative configuration of molecules rather complicated in general~\cite{TMori84BCSJ,TMori98BCSJ}. 
In contrast, $V_{ij}$ obeys more or less a monotonic function of the distance 
because it is the Coulomb interaction. 
Therefore the interchain part of $V_{ij}$ is not necessarily small  
compared to the intrachain one even in these quasi-1D systems~\cite{Fritsch91JPIF,Castet96JPIF}. 
However, let us neglect the interchain interactions first 
and discuss this issue later in $\S$~\ref{subsec_interchain} and \ref{subsec_frustration}. 

Estimations of the Coulomb energies in these molecular 
systems are difficult at present. Quantum chemistry calculations are performed for an isolated 
molecule or clusters of them, which provide unrealisticly large values 
since the screening effect in solids is left out, 
while such estimations from the first principle in these molecular solids 
are still yet to be done. 
However it is believed to be of the order of $U \simeq 1$ eV 
from different measurements and estimates, 
which gives $U/|t_\textrm{intra}| \simeq 5$ for DCNQI$_2X$ and TMTTF$_2X$,  
while for TMTSF$_2X$ smaller values of $U/|t_\textrm{intra}| \simeq 3$. 
The estimated values for the intrachain Coulomb energy $V_\textrm{intra}$ 
are again ambiguous but many provide rather large values: 
the ratio $V_\textrm{intra}/U$ 
in a range about $0.2 \sim 0.6$~\cite{Mila95PRB,Fritsch91JPIF,Castet96JPIF,Imamura98CPL}. 

A crucial difference between these two families is that the stacking of DCNQI molecules is uniform 
while that of TM moluecules is slightly dimerized, 
as seen in Fig.~\ref{fig_DCandTM}.
Thus a minimal effective model to investigate 
the occurance of CO in these compounds is the 1D dimerized EHM, represented as, 
\begin{eqnarray} 
{\cal H}_{\rm 1D} =-t \sum_{i\sigma}  (1+(-1)^i \delta_\textrm{d}) (c_{i+1\sigma}^\dagger c_{i\sigma} + h.c.) \nonumber \\
 + U \sum_{i} n_{i\uparrow} n_{i\downarrow} + V \sum_{i} n_i n_{i+1}, 
\label{eqn:extHub1D} 
\end{eqnarray}
where $i$ is the site index along the chain. 
Here, the transfer integrals allow dimerization as alternating $t(1+\delta_\textrm{d})$ and $t(1-\delta_\textrm{d})$; 
in the DCNQI compounds $\delta_\textrm{d}=0$, 
while for the TM compounds values of $t_{a1}$ and $t_{a2}$~\cite{TMori82CL,Grant83JDP,Ducasse86JPC,TMori84BCSJ} 
(for the indices see Fig.~\ref{fig_DCandTM}) read $\delta_\textrm{d} \lesssim 0.1$. 
The inter-site Coulomb repulsion between neighboring sites is set to be uniform as $V$, 
which is an approximation for TM$_2X$. 
Again, the value of $\delta_\textrm{d}$ does not directly result in a similar value of dimerization in $V_\textrm{intra}$. 
In fact, quantum chemistry calculations for TM$_2X$~\cite{Fritsch91JPIF,Castet96JPIF} 
provide the dimerization parameters in $V_{a1}$ and $V_{a2}$ to be less than 1~\%. 
For the DCNQI compounds the electron filling is one quarter while it is three quarter for the TM compounds, 
but in this 1D model the two situations are equivalent since 
electron-hole symmetry holds. Note that other effects may break this symmetry then 
we should treat the two cases separately.

\begin{figure}
\centerline{\includegraphics[width=5truecm]{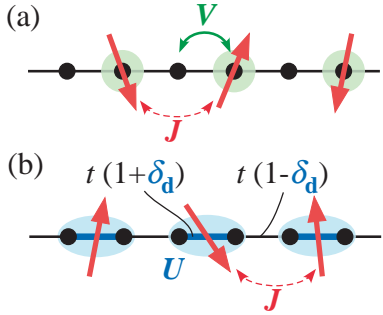}}
\vspace*{-1em}
\caption{(Color online) 
Two limiting cases of strongly correlated insulators 
in quarter-filled system~\cite{Seo04CR,Fukuyama00Physica}~: 
(a) the charge ordered insulating state and (b) the dimer Mott insulating state. 
Black dots and colored area represent the lattice sites and 
the localized carriers, respectively, 
while the thickness of the bonds show the difference in the 
transfer integrals.  
The arrows show the paramagnetic localized spins.  
}
\vspace*{-1.5em}
\label{fig_COvsDM}
\end{figure}

It is useful to describe two limiting cases for the insulating states 
in this model at quarter-filling due to strong Coulomb interaction~\cite{Seo04CR,Fukuyama00Physica}, 
which are shown in Fig.~\ref{fig_COvsDM}. 
These states, in a broad sense, are realized 
in a wide range of $A_2B$ compounds, not only in quasi-1D but also in quasi-2D systems discussed in $\S$~\ref{sec_q2d}. 
One is the Wigner crystal-type CO state stabilized by $V$ in the presence of strong $U$, 
and the other is the Mott insulating state stabilized by $U$ in the presence of strong dimerization $\delta_\textrm{d}$: 
a dimer Mott insulator. 
The spin degree of freedom in such insulating states would behave as 
an $S=1/2$ localized spin chain, where the spins are located 
on every other site for the CO state whereas on every dimer for the dimer Mott insulating state. 
Roughly speaking, charge localization is determined by large energy scales 
such as $U$ and $V$, while spin properties are determined by 
a smaller energy scale of the order of the Heisenberg coupling, $J$, 
acting between these localized spins. 
We will discuss their detailed properties in the following subsections. 

\begin{figure}
\centerline{\includegraphics[width=5.8truecm]{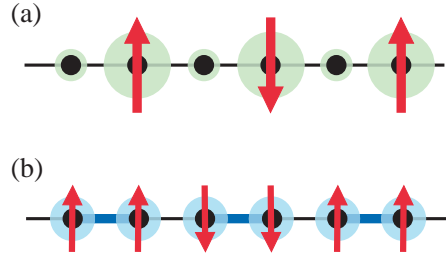}}
\vspace*{-1em}
\caption{(Color online) 
Typical mean-field solutions in one-dimensional dimerized 
extended Hubbard model at quarter-filling~\cite{Seo97JPSJ}, 
${\cal H}_{\rm 1D}$ in eq.~(\ref{eqn:extHub1D}): 
(a) the charge ordered antiferromagnetic insulating state for $\delta_\textrm{d}=0$ and large $U$ and $V$, and,  
(b) the dimer-type antiferromagnetic insulating state for $\delta_\textrm{d} \neq 0$, large $U$, and $V=0$. 
The size of the colored circles and the arrows represents the charge density 
and the amount of spin moment on each site. 
The charge densities are (a) disproportionated alternatively as,  
$1/2+\delta$, $1/2-\delta$, $1/2+\delta$, $1/2-\delta$, 
where $\delta$ is the amount of charge disproportionation, 
and (b) uniform at 1/2. 
}
\vspace*{-1.5em}
\label{fig_1dMF}
\end{figure}

The MF solutions mentioned above 
are consistent with such two limiting cases. 
This is seen in the obtained spin and charge patterns 
for the two cases, schematically shown in Fig.~\ref{fig_1dMF}, 
and actually many works have been performed on ${\cal H}_{\rm 1D}$ 
based on such MF results~\cite{Kishigi00JPSJ}. 
However we should keep in mind that this MF treatment cannot correctly 
describe such insulators at strongly correlated regime in general, 
for example the paramagnetic insulating phase at temperatures above the magnetically ordered phases 
observed in experiments cannot be reproduced. 
Moreover in purely 1D models the role of quantum fluctuations which is left out in MF 
is crucial, and we will see that it considerably modifies the MF results. 

We note that the cations/anions are at crystallographically 
equivalent positions from the DCNQI/TM molecules 
at high temperatures so that, essentially, they do not contribute to 
the electronic properties. 
However, in the case of non-centrosymmetrical anions in TM$_2X$, 
the anions show a disorder-to-order transition by lowering temperature 
and generate a potential with longer periodicity than the original unit cell, 
resulting in modifications in the one-particle properties~\cite{Ravy04CR}. 
We do not discuss such cases in this paper since its relation 
with the CO due to electron correlation is still obscure, 
but the fact that the anion ordering produces charge density modulations 
suggests the importance of the anions in some cases. 
Actually possible roles of the $X$ unit in stabilizing CO states 
through its coupling to the electron system will be addressed in $\S$~\ref{subsec_lattice}. 

\subsection{One-dimensional extended Hubbard model}\label{subsec_1dEHM}

\vspace{3pt}

The quarter-filled 1D EHM, ${\cal H}_{\rm 1D}$ in eq.~(\ref{eqn:extHub1D}) with $\delta_\textrm{d}$=$0$, 
was first considered by Ovchinnikov~\cite{Ovchinnikov73Sov} who 
discussed that in the $U/t=\infty$ limit 
a CO insulating ground state is realized for $V>2t$ $(=V_\textrm{cr})$, 
as in the following. 
In this limit, double occupancy is strictly prohibited 
so that every site is either occupied or unoccupied. 
Furthermore the spin degree of freedom is frozen 
since the ``order" of electrons cannot be changed due to the 1D nature. 
Then the charge degree of freedom is equivalent to  
a spinless fermion (SF) model at half-filling with the nearest neighbor Coulomb interaction $V$, i.e., 
the interacting SF model, and the spin degree of freedom acts freely as Curie spins. 

This SF model, by identifying occupied site as up spin and unoccupied site as down spin, 
can be mapped onto  a 1D $S=1/2$ (pseudo-)spin model through Jordan-Wigner transformation. 
Quarter-filling in the original electronic system 
corresponds to half-filling in the SF model 
where half of the sites would be occupied,  
and to total magnetic moment of zero in the effective spin model.
The inter-site Coulomb interaction between SF transforms to an antiferromagnetic 
interaction, $J_z S_i^z S_j^z$ with $J_{z}=V$, 
while the kinetic enery term becomes as 
$J_{xy} ( S_i^x S_j^x + S_i^y S_j^y )$ with $J_{xy}=2t$. 
This is an XXZ model, 
where a $T=0$ phase transition from a gapless ``XY" state ($J_z < J_{xy}$) 
to an antiferromagnetic ``Ising" state ($J_z > J_{xy}$)  is known~\cite{Yang66PR}. 
Transforming back to the SF model, these correspond to 
a metallic Tomonaga-Luttinger liquid (TLL) and the CO insulating state, repectively; 
the critical point is the Heisenberg point $J_{xy}=J_z$, 
i.e., $V=V_\textrm{cr}$. 

It is noteworthy that this is in fact a rigorous demonstration of a ``classical" picture of CO 
discussed in systems regardless of the dimension~\cite{Anderson56PR}. 
In general, the CO problem, in the condition of 
kinetic energy set to 0 ($t_{ij}=0$) in addition to infinite $U$ with no double occupancy discussed above,  
can be described by SF's interacting only by the inter-site Coulomb repulsion $V_{ij}$. 
This can be mapped onto Ising models, $J_{ij} \sigma_i \sigma_j$, 
with $J_{ij} = V_{ij}$, similarly to the $J_z$-term in the 1D case above. 
Therefore, analogies between spin systems and CO systems can be generally expected 
in the presence of strong correlation, 
although the mapping is only approximate for 2D and 3D systems  
with finite $t_{ij}$ and $U$, and even for 1D systems with finite $U$. 

\begin{figure}
\centerline{\includegraphics[width=7.0truecm]{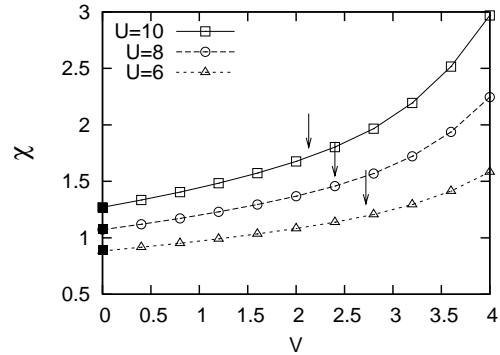}}
\vspace*{-1em}
\caption{Spin susceptibility $\chi$ calculated by exact diagonalization 
as a function of {\it V} for several values
of {\it U} at quarter filling ($t=1$)~\cite{Tanaka05JPSJ}. 
Solid squares represent exact results in
the one-dimensional Hubbard model and arrows indicate the points 
above which the charge gap opens.
[After ref.~\ref{Tanaka05JPSJ}.]
}
\vspace*{-1.5em}
\label{fig:tanaka}
\end{figure}

Since the charge and spin degrees of freedoms are completely decoupled in 
the $U/t=\infty$ limit, the ground state wave function has the form 
$| \Phi \rangle = | \phi_{\rm SF} \rangle | \chi_\sigma \rangle$, 
where $| \phi_{\rm SF} \rangle$ is the ground state of SF's and 
$| \chi_\sigma \rangle$ denotes the spin configuration. 
On the other hand, at finite $U/t$, these spins interact with each other. 
In the large-$U$ region, recently 
Tanaka and Ogata~\cite{Tanaka05JPSJ} analytically discussed 
the $V$-dependence of the magnetic susceptibility $\chi$.
This is done by extending the large-$U$ studies of the 1D Hubbard model 
based on the Bethe ansatz~\cite{Ogata90PRB,Shiba91IJMPB}, 
where the charge degree of freedom can be represented by SF's, 
and the spin degree of freedom is described by Heisenberg spin chains 
with an effective exchange coupling $J_{\rm eff}$. 
For the 1D EHM, according to the degenerate perturbation theory~\cite{Shiba91IJMPB}, 
the degeneracy of the spin system is lifted by a term proportional to $J=4t^2/U$. 
However, because of the presence of the charge degree of freedom, 
$J_{\rm eff}$ is not just equal to $J$, 
but also depends on the probability of finding two 
SF's on nearest-neighbor sites, namely, 
$\langle n_i n_{i+1}\rangle_{\rm SF} \equiv \langle \phi_{\rm SF} | n_i n_{i+1} | \phi_{\rm SF} \rangle$. 
(There is another contribution to $J_{\rm eff}$ discussed in ref.~\ref{Tanaka05JPSJ}.) 
This value decreases as $V$ increases
and its derivative with respect to $V$ is continuous even at $V_\textrm{cr}/t$~\cite{Yang66PR},
where the CO transition occurs. This is because the phase transition is a
Berezinskii-Kosterlitz-Thouless (BKT) type\cite{Berezinskii71JETP,Kosterlitz73JPC}, 
which is essentially different 
from either first or second order transition.
The divergence of the correlation length is more rapid than any power 
law, and then the gap increases very slowly near the critical point: 
the charge gap behaves as $\Delta_{\rm c} \simeq 4\pi
\exp(-\pi^{2}/2(2(\rho-1))^{\frac{1}{2}})$,~\cite{Tanaka05JPSJ,desCloizeaux66JMP} with 
$\rho = V/2t \geq 1$. 
Reflecting these, $\chi$ increases as a monotonic function of $V$ 
since $\chi$ is proportional to $1/J_{\rm eff}$~\cite{Ogata90PRB,Shiba91IJMPB}, 
continiously through the CO transition point (see also Fig.~\ref{fig:tanaka}).
This indicates that the fluctuation is large near the phase boundary even in
the CO states, and consequently CO has small effect on 
the spin degree of freedom.

\begin{figure}
\centerline{\includegraphics[width=6.0truecm]{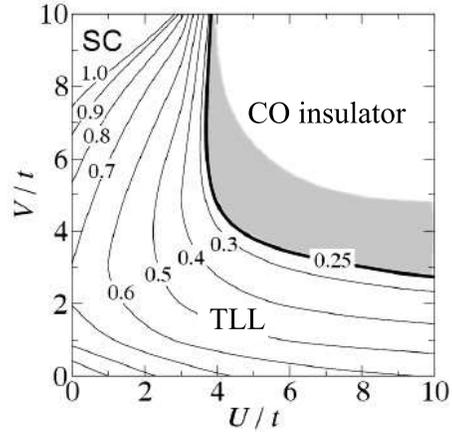}}
\vspace*{-1em}
\caption{
Ground state phase diagram of the one-dimensional extended Hubbard model 
at quarter-filling on the plane of $U/t$ and $V/t$~\cite{Ejima05EPL}.  
CO stands for charge order and TLL for Tomonaga-Luttinger liquid. 
SC represents the superconducting phase. 
Contour map for the Tomonaga-Luttinger parameter $K_\rho$ is shown. 
The bold line represents the boundary for the metal-insulator transition. 
The shaded area is the region with an exponentially small gap.
[By courtesy of S. Ejima.]
}
\vspace*{-1.5em}
\label{fig_1dEHM}
\end{figure}

For whole finite values of $U$ and $V$, 
Mila and Zotos~\cite{Mila93EPL} first provided 
the ground state phase diagram on the plane of $U/t$ and $V/t$, 
using numerical Lanczos exact diagonalization (ED) method~\cite{Dagotto94RMP} 
for finite size systems up to $L=16$ sites.  
They have judged the metal-insulator phase boundary 
by calculating the charge gap for finite systems as  
$\Delta_{\rm c}=[ E_0(L/2+1)+E_0(L/2-1)-2E_0(L/2) ]/2$, 
where $E_0(n)$ is the ground state energy for $n$ electron system 
($n=L/2$ for quarter-filling), 
and extrapolating the results to the thermodynamic limit. 
Later, different numerical methods have been applied 
by different authors~\cite{Penc94PRB,Sano94JPSJ,Nakamura99PRB,Clay03PRB,Ejima05EPL}~ 
and the main features are confirmed; 
the phase diagram by Ejima \et~\cite{Ejima05EPL} is shown in Fig.~\ref{fig_1dEHM}, 
which is determined by the density matrix renormalization group (DMRG) method 
treating system sizes up to more than $L=100$ sites. 
The CO insulating state is in the large $(U/t,V/t)$ region, 
while the metallic state in the rest is the TLL. 
The TLL parameter $K_\rho$, 
which is the exponent characterizing the power of the correlation function, 
becomes equal to $K_\rho=1/4$ at the phase boundary 
($K_\rho=1$ at $U=V=0$). 
As $U$ is increased, this phase boundary 
approaches to $(U/t=\infty,V/t=2)$, the exact value explained above.

The magnetic susceptibility $\chi$ has been discussed 
in ref.~\ref{Tanaka05JPSJ} for such finite $(U/t,V/t)$. 
This is done numerically by the Lanczos ED method up to $L=16$ sites, 
as shown in Fig. \ref{fig:tanaka} where $\chi$ at $T=0$ for several values of
$U/t$ and $V/t$ are plotted.
The TLL property $\chi=2/\pi u_\sigma$ 
is used with $u_\sigma$ being the velocity of the spin excitation 
(i.e., the spin velocity)~\cite{Ogata91PRL}.
The critical values for the metal-CO insulator transition 
are indicated by arrows in Fig.~\ref{fig:tanaka}.
One can see that $\chi$ increases with $V$, 
namely, $J_{\rm eff}$ is suppressed as in the large-$U$ case above. 
$\chi$ smoothly varies as a function of $V$ even 
through the CO phase transition, as in the exact calculation 
in the large-$U$ case~\cite{Tanaka05JPSJ}. 
The spin susceptibility is again not affected much by the opening of a charge gap. 
This is consistent with field theoretical studies 
that we will see in $\S$~\ref{subsec_boson}, 
indicating the transition to be of the BKT type for finite $(U/t,V/t)$ as well. 

\begin{figure}
\centerline{\includegraphics[width=6truecm]{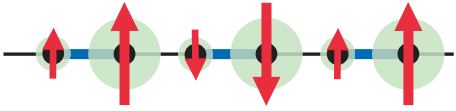}}
\vspace*{-1em}
\caption{(Color online) 
Coexistence of charge order and dimerization in the one-dimensional dimerized 
extended Hubbard model for TM$_2X$. 
This is a schematic drawing of the mean-field solution, where 
antiferromagnetism also coexists~\protect\cite{Seo97JPSJ}. 
Charge disproportionation exists as in the charge ordered state in Fig.~\ref{fig_1dMF}~(a)
whereas the amount of spin moment, parallel whithin dimers and antiparallel between them, 
are also disproportionated in a similar way. 
} 
\vspace*{-1.5em}
\label{fig_1dMF2}
\end{figure}

In the presence of finite dimerization $\delta_\textrm{d}$, 
it has been well known for the $V=0$ case, namely, the 1D dimerized Hubbard model, that 
$\delta_\textrm{d}$ together with $U$ always 
give rise to the dimer Mott insulating state  
(Fig.~\ref{fig_COvsDM}(b)) for quarter-filling. 
This is because $\delta_\textrm{d}$ opens an energy gap at $\pm 2\kf$ 
in the non-interacting energy dispersion 
making the lower band effectively half-filled. 
Then even for an infinitesimal repulsion 
the system becomes a Mott insulator which is characteristic of 
1D half-filled electronic systems. 
This situation have been theoretically investigated extensively~\cite{Mila01Review}, 
from the large-$U$ case~\cite{Bernasconi75PRB,Penc94PRB_2} to the small-$U$ case~\cite{Emery82PRL}.
When finite $V$ is added, 
there arises a competition between two insulating states, 
this dimer Mott insulator and the CO insulator.~\cite{Seo97JPSJ} 
The CO state here can be considered as a coexistence of 
dimerization and CO as seen in Fig.~\ref{fig_1dMF2}. 
Accurate numerical calculations are yet to be done 
and now in progress,~\cite{Mila95PRB,Nishimoto00JPSJ,Shibata01PRB,Otsuka04PRB}
so let us discuss such competition in the next subsection.

\subsection{Bosonization}\label{subsec_boson}

\vspace{3pt}

The bosonization method is one of the most powerful tools 
to treat 1D systems analytically~\cite{1DRecentReview}.   
In this approach, 
quantum fluctuation, which plays an essential role on the
electronic state in 1D systems, can be fully taken into account, 
in contrast to the MF theory. 
Furthermore, the analytical treatment enables us 
to capture insight of the physics, e.g., 
found in numerical calculations discussed in the previous subsection. 
It has been successfully applied to ${\cal H}_{\rm 1D}$ in eq.~(\ref{eqn:extHub1D}) 
by Yoshioka \et~\cite{Yoshioka00JPSJ,Yoshioka01JPSJ}, 
as in the following, 
who clarified the relation between such lattice models 
and the conventional ``$g$-ology" picture. 

In the ordinary bosonization procedure, 
only the one-particle fermion states around the two Fermi wavenumber, $\pm \kf$,
are considered and the four body interaction in the Hamiltonian is 
expressed in terms of these states. 
In quarter-filled systems, 
the 8$\kf$-Umklapp scattering
exists\cite{Giamarchi92PRB,Schulz94Book,Giamarchi97PhysicaB,Yonemitsu97PRB}
because of $\kf=\pi/(4a)$ with $a$ being
the lattice spacing and it is crucial for 
the appearance of the insulating state.   
However, one can not obtain this Umklapp scattering by the ordinary method above. 
This is because it can only be expressed by the interaction
processes where four right-going electrons are scattered 
into the left-going states and vice versa to gain (or loose) 8$\kf$, 
and then the scattering should include higher order interaction processes   
through the one-particle state around $\pm 3 \kf$.

\begin{figure}
\centerline{\includegraphics[width=8truecm]{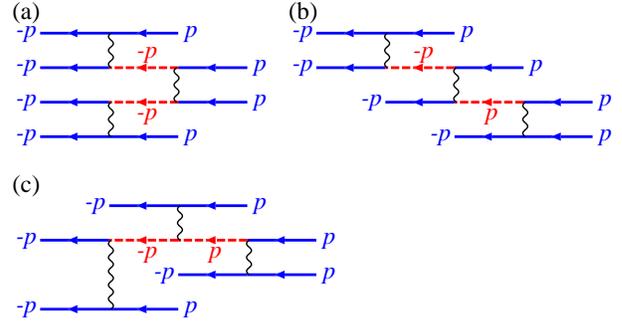}}
\vspace*{-1em}
\caption{(Color online) 
Diagrams representing the $8\kf$-Umklapp scattering. 
Here $p=+/-$ expresses the right/left-going one-particle states 
and the solid and the dotted lines express the electrons near $\pm \kf$  and
 $\pm 3 \kf$, respectively. 
}
\vspace*{-1.5em}
\label{fig:Umklapp}
\end{figure}

A systematic way to derive the $8\kf$-Umkpapp scattering 
has been developped~\cite{Yoshioka00JPSJ,Yoshioka01JPSJ} 
as follows. 
The one-particle states are divided into two parts: 
the states near $\pm \kf$ and those near $\pm 3 \kf$. 
The effective Hamiltonian written in terms of the former states 
near $\pm \kf$ is obtained by integrating out the one-particle states
near $\pm 3\kf$ and treating the interaction processes including 
both the states near $\pm \kf$ and those near $\pm 3 \kf$ perturbatively.    
The $8\kf$-Umklapp scattering appear in the third order interaction 
processes shown in Fig. \ref{fig:Umklapp}. 

Then we can express the effective phase Hamiltonian for 
the 1D EHM with $\delta_\textrm{d}=0$, 
by using the bosonic phase variables 
representing the charge and spin fluctuations~\cite{FukuyamaTakayama}. 
The Hamiltonian is divided into the charge part, ${\cal H}_{\rho}$, 
and the spin part, ${\cal H}_{\sigma}$, 
which is essentially the same as that of the 1D isotropic 
Heisenberg model with a gapless excitation. 
${\cal H}_\rho$ is given by the sine-Gordon model shown below,
\begin{align}
 {\cal H}_\rho =& \frac{v_\rho}{4 \pi} \int \d x
\left\{
\frac{1}{K_\rho} (\delx \theta_\rho)^2 + 
K_\rho (\delx \phi_\rho)^2
\right\} \nonumber \\
&+ \frac{g_{1/4}}{2 (\pi \alpha)^2} \int \d x \cos 4 \theta_\rho, 
\label{eqn:boson}
\end{align}
where $[\theta_\rho(x), - \partial_{x'} \phi_\rho(x')/(2 \pi)] = \im
\delta (x-x')$, and
$v_\rho$ and $K_\rho$ are the velocity of the charge excitation and 
the TLL parameter introduced in $\S$~\ref{subsec_1dEHM},
respectively. 
The quantity $\alpha^{-1}$ is the ultraviolet cutoff of the order of
$a^{-1}$. 

The coefficient of the non-linear term $g_{1/4}$ expressing 
the $8\kf$-Umklapp scattering is written as,
$g_{1/4} = (Ua)^2(Ua - 4Va) / 8 (\pi t \alpha)^2$.
The order parameter of CO at the $i$-th site, $O_{\rm CO}(x_i)$ 
is given by using the phase variables, as 
$O_{\rm CO}(x_i) \propto \cos (4\kf x_i + 2 \theta_\rho) = (-1)^i \cos 2 \theta_\rho$,  
where $x_i = i a$. 
This indicates that 
CO is stabilized for $\theta_\rho = 0$ and $\theta_\rho =
\pi/2$, whereas it disappears for $\theta_\rho = \pi/4$ and $\theta_\rho
= 3 \pi /4$. 
The former and the latter can be realized in the case of $g_{1/4} < 0$ and $g_{1/4} > 0$, respectively. 
This shows explicitly the fact that 
the inter-site Coulomb repulsion $V$ stabilizes the CO state
through the decrease of $g_{1/4} < 0$ as $V$ is increased. 
We note that this condition $g_{1/4} < 0$ for the appearance of CO, i.e., $4V > U$, 
is also a necessary condition in the MF treatment. 
This is based on the charge susceptibility within random phase approximation (RPA) , 
$\chi_{\rm c}(q) = {\chi_0(q)}/[ {1 + (U + 4 V \cos qa)\chi_0(q)}]$, 
where $\chi_0(q)$ is the noninteracting charge susceptibility taking a positive value; 
$4V > U$ is required for $\chi_{\rm c}$ to diverge at $qa=\pi$, 
i.e., for the realization of the CO state with $q=4\kf$. 

\begin{figure}
\centerline{\includegraphics[width=6.5truecm]{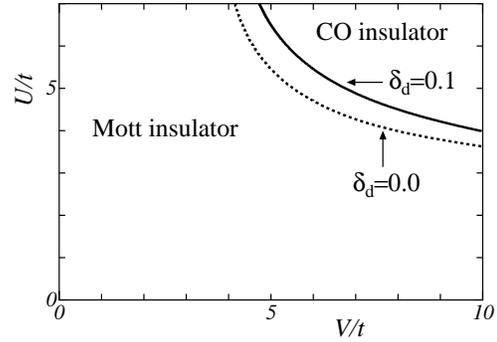}}
\vspace*{-1em}
\caption{
Ground state phase diagram 
of the one-dimensional dimerized extended Hubbard model at quarter-filling 
on the plane of $V/t$ and $U/t$, 
determined by the renormalization group equations for 
the bosonized phase Hamiltonian. 
The solid curve expresses the phase boundary between 
the charge ordered (CO) insulator and the dimer-type Mott insulator 
for $\delta_\textrm{d} = 0.1$~\cite{Tsuchiizu01JPSJ}, 
while the dotted curve shows the boundary for $\delta_\textrm{d} = 0.0$
where the Mott insulating phase is replaced 
by the Tomonaga-Luttinger liquid~\cite{Yoshioka00JPSJ} (see Fig.~\ref{fig_1dEHM}). 
}
\vspace*{-1.5em}
\label{fig:dimerization_phase}
\end{figure}

The low energy property of the phase Hamiltonian ${\cal H}_\rho$ 
is determined by the renormalization group equations
by introducing $l=\ln(\alpha'/\alpha)$ 
with a new scale $\alpha'$ ($>\alpha$), 
\begin{align}
 \frac{\d}{\d l} K_\rho(l) &= - 8 G_{1/4}(l)^2 K_\rho(l)^2, \\
 \frac{\d}{\d l} G_{1/4}(l) &= (2 - 8  K_\rho(l))G_{1/4}(l). 
\end{align} 
The initial conditions are 
$K_\rho(0) = K_\rho$ and $G_{1/4}(0) = g_{1/4}/(2 \pi v_\rho)$. 
Whether the system is in the metallic TLL state or in the CO insulating state 
is determined by the $l \rightarrow \infty$ solutions
characterized by $G_{1/4}(\infty) =0$ or $G_{1/4}(\infty) = -\infty$, 
respectively. 
The phase boundary determined by the renormalization group equations, 
where $K_\rho(\infty) = 1/4$, is shown by the
dotted curve in Fig. \ref{fig:dimerization_phase} where $\alpha=2a/\pi$ is used. 
This phase transition is of BKT type 
since the sine-Gordon theory is equivalent 
to the low-energy properties of the classical 2D XY model.  
The phase diagram obtained by the present bosonization theory 
is qualitatively the same as that shown 
in Fig. \ref{fig_1dEHM}. 

The bosonization theory above can be straightforwardly extended 
to include the dimerization $\delta_\textrm{d}$ together with finite $V$~\cite{Tsuchiizu01JPSJ}. 
As a result of the energy gap at $\pm 2\kf$ 
in the non-interacting energy dispersion 
due to finite value of $\delta_\textrm{d}$, 
in addition to ${\cal H}_{\rho}$ in eq.~(\ref{eqn:boson}),  
the $4\kf$-Umklapp scattering is generated.~\cite{Emery82PRL} 
This is expressed in terms of the phase variables as, 
\begin{align}
 {\cal H}_{1/2} &= - \frac{g_{1/2}}{2 (\pi \alpha)^2}
\int \d x \sin 2 \theta_\rho, \label{eqn:4kfUmklapp}
\end{align}
where $g_{1/2}$ is proportional to $\delta_\textrm{d}$ for $|\delta_\textrm{d}| \ll 1$.
The ground state phase diagram in the presence of $\delta_\textrm{d}$
determined by the renormalization group equations, 
on the plane of $V/t$ and $U/t$, is shown in Fig.~\ref{fig:dimerization_phase}. 
Due to the $4\kf$-Umklapp scattering, 
the metallic TLL state changes to the dimer Mott insulating state 
with uniform charge distribution even for infinitesimal interactions, 
as expected from the works for $V=0$~\cite{Mila01Review}. 
The CO insulating state (Fig.~\ref{fig_1dMF2}) is suppressed because the
commensurability energies due to the two kinds of Umklapp scattering
compete with each other for $g_{1/4}<0$ and 
${\cal H}_{1/2}$ has a larger scaling dimension. 
Nevertheless, importantly, the CO phase still exists in the large $(U/t,V/t)$ region.     
This model for the charge sector having two non-linear terms 
is called the double frequency sine-Gordon model 
and it is suggested that the phase boundary 
between the dimer Mott insulator and the CO insulator 
is an Ising type phase transition, 
where the charge gap $\Delta_{\rm c}$ becomes 0 at the phase boundary~\cite{Tsuchiizu02JPCS,Otsuka04PRB}. 

\subsection{Interchain coupling}\label{subsec_interchain}

\vspace{3pt}

In purely 1D systems, 
phase transitions do not occur at finite temperatures 
because any ordered state collapses due to thermal fluctuations.  
In this sense, the critical temperature for CO, $T_{\rm CO}$ 
in the 1D EHM that we have discussed in $\S$~\ref{subsec_1dEHM} and~\ref{subsec_boson} 
is always absolute zero temperature. 
However, quasi-1D materials such as DCNQI$_2X$ and TM$_2X$ 
undergo phase transitions at finite temperatures,  
which is a consequence of the three (or two) dimensionality. 
In the electronic sector, there exist two kinds of interchain couplings; 
one is the single particle hopping between the chains ($t_\textrm{inter}$) and 
the other is the mutual Coulomb interaction between electrons in different chains ($V_\textrm{inter}$). 
As mentioned in $\S$~\ref{subsec_COq1d}, 
$t_\textrm{intra}/t_\textrm{inter}$ is of the order of 10 in the DCNQI  and TM salts, 
while the interchain $V_\textrm{inter}$ may be of the same order with the intrachain $V_\textrm{intra}$. 
There are quantum chemistry studies for TM$_2X$ actually 
showing that it the case~\cite{Fritsch91JPIF,Castet96JPIF}. 

The interchain hopping $t_\textrm{inter}$ generates  
warping of the Fermi surface in the non-interacting band structure. 
In TMTSF$_2X$ compounds, because of this effect, 
together with the weaker effect of Coulomb repulsion than in their TMTTF analogs, 
the strongly correlated insulators such as CO and dimer Mott insulators are not realized. 
Then the Fermi surface persists down to low temperatures. 
Such a dimensional crossover in half-filled models are theoretically discussed~\cite{Kishine02IJMPB} 
while that in the quarter-filled CO systems remains to be investigated. 
This should be relevant to the pressure-induced melting of CO observed in DI-DCNQI$_2$Ag~\cite{Itou04PRL}, 
where the CO transition turns from a second order to a first order phase transtition by applying pressure, 
and a region with $T^3$-resistivity appears in the metallic state at the verge of the CO phase. 
In TM$_2X$, in contrast, 
the system becomes unstable due to the nesting of the Fermi surface 
toward a metal-insulator transition to 
an incommensurate spin-density-wave (SDW) ground state, 
which have also been studied intensively~\cite{Ishiguro98Book,Kagoshima89Book,Jerome82AdvPhys}. 

Effects of the interchain Coulomb repulsion on the CO state 
have been discussed recently by Yoshioka \et\cite{YTS05} 
by using the bosonization method described in the previous subsection. 
When one considers only one kind of uniform coupling in the interchain direction, it is expressed as,
\begin{align}
 {\cal H}_\bot &= V_\bot \sum_{i,l} n_{i,l} n_{i,l+1}, 
\end{align} 
where $l$ denotes the chain index. 
This may be applied to the DCNQI compounds as the interchain coupling is uniform and isotropic 
while in TM$_2X$ the network is more complicated as seen in Fig.~\ref{fig_DCandTM} 
(we will consider them in $\S$~\ref{subsec_frustration}). 
In the ``interchain MF" treatment,~\cite{Scalapino75} 
the interchain part is treated within MF approximation as, 
$n_{i,l} n_{i,l+1} \to \langle n_{i,l} \rangle n_{i,l+1} + n_{i,l}
\langle n_{i,l+1} \rangle - \langle n_{i,l} \rangle \langle n_{i,l+1} \rangle $, 
while quantum fluctuation within the 1D chains is incorporated.  
When $\langle n_{i} \rangle$ along the chains is written as $1/2 + (-1)^{i}
\delta$ where $\delta$ being the amplitude of CO, 
and an antiphase pattern of the Wigner crystal-type CO is assumed, 
the problem is reduced to an effective 1D model where $\delta$ is 
determined self-consistently. 
If $\delta$ is finite, 
an energy gap opens at $\pm 2 \kf$ and as a result 
the 4$\kf$-Umklapp scattering, $\tilde{{\cal H}}_{1/2}$, is generated, 
similarly to the case with finite dimerization discussed in $\S$~\ref{subsec_boson}. 
However this Umklapp scattering is written with a cosine function of the phase variable 
but not with a sine function as in eq.~(\ref{eqn:4kfUmklapp}), as follows, 
\begin{align} 
 \tilde{{\cal H}}_{1/2} &= \frac{\tilde{g}_{1/2}}{2(\pi \alpha)^2} 
\int \d x \cos 2 \theta_\rho, 
\end{align} 
where $\tilde{g}_{1/2}$ is proportinal to $\delta$ as long as $|\delta| \ll 1$. 
Therefore, different from the case with dimerization the generated non-linear term does not compete 
but cooperate with the $\cos 4 \theta_\rho$ term when $g_{1/4} < 0$ in eq.~(\ref{eqn:boson}). 
We note that such term, which acts as an alternating on-site potential upon the 1D chain, 
has been discussed from another context~\cite{Penc94PRB_2}. 
Thus, the interchain interaction stabilizes the antiphase CO, 
and the critical temperature $T_{\rm CO}$ becomes finite.   
Renormalization group study on this phase Hamiltonian~\cite{YTS05}  
indicates that an infinitesimal value of $V_\bot$ gives rise to finite $T_{\rm CO}$ 
as long as $K_\rho<1/2$ in the original 1D EHM model. 
This results in a sudden enlargement of the CO insulating region in the $U-V$ phase diagram
once $V_\bot$ is considered,  
from the case of the 1D EHM where the phase boundary is determined by $K_\rho = 1/4$, 
as in Fig.~\ref{fig_1dEHM}. 

\subsection{Coupling to the lattice}\label{subsec_lattice}

\vspace{3pt}

1D electron systems are unstable when 
there exist couplings to the lattice degree of freedom. 
A well known case is the 2$\kf$ CDW (Peierls) instability, 
where electron gas becomes unstable at low temperatures 
toward a lattice distortion of period 2$\kf$ 
when the nesting condition of the Fermi surface is satisfied~\cite{Kagoshima89Book}. 
In the purely 1D case this condition is perfect 
so that any 1D electron system is susceptible to the Peierls-Fr\"{o}lich state. 
Another case is when the electrons are localized due to strong correlation 
and the spin degree of freedom is described by a 1D Heisenberg chain. 
There, the spin-lattice coupling makes the system unstable toward 
a spin-singlet formation accompanied with a lattice distortion, i.e., 
the spin-Peierls (SP) transition~\cite{Bray83Book,Fukuyama86Book}. 
In fact, SP transition is frequently observed 
in many quasi-1D $A_2B$ compounds, while its interplay 
with the charge sector such as the CO and the dimer Mott insulators is not obvious. 

Let us first consider the so-called Peierls-type coupling 
to the lattice degree of freedom, treated as classical variables here
neglecting quantum fluctuations, 
in addition to ${\cal H}_{\rm 1D}$ in eq.~(\ref{eqn:extHub1D}) with $\delta_\textrm{d} = 0$. 
This allows modulations of the transfer integrals 
at the cost of the elastic energy, i.e., 
$t \rightarrow t(1+u_{i,i+1}) + K u_{i,i+1}^2 /2$, 
where $u_{i,i+1}$ is the dimensionless modulation of the lattice, 
controlled by the actual movement of the molecules $i$ and $i+1$.  
$K$ is the renormalized lattice constant, which is taken to be uniform here 
but in general, as in the discussion in $\S$~\ref{subsec_COq1d} on $V_{ij}$, can be modulated; 
for example in the 1D model of TM$_2X$ it would be dimerized 
as $K_{a1}$ and $K_{a2}$ (see Fig.~\ref{fig_DCandTM}). 
Hereafter we write $u_{i,i+1} = u_i$ for simplicity.
The ``softness" of the lattice is indicated by $1/K$; 
at $K \rightarrow \infty$ the system is in the ``hard" limit 
so that the lattice does not move and then 
the model reduces to the purely electronic model, ${\cal H}_{\rm 1D}$.

One serious effect of this coupling 
is that it can generate the dimer Mott insulating state 
even without intrinsic dimerization ($\delta_\textrm{d}=0$). 
This is understood in the $U/t=\infty$ limit discussed by 
Bernasconi \et~\cite{Bernasconi75PRB}. 
Extending the discussion in $\S$~\ref{subsec_1dEHM}, 
the charge degree of freedom described by 
the half-filled interacting SF or equivalently the $S=1/2$ XXZ model 
is now coupled to the lattice. 
In the latter model the Peierls coupling appears in the XY term, $J_{xy}$; 
for $J_{xy}=2t  \ge J_{z}=V$, infinitesimal spin lattice coupling $1/K$ 
can drive the ground state toward SP lattice dimerization~\cite{Bray83Book}.  
This is actually the dimer Mott insulator when one transforms back to the original electron model. 
This state competes with the CO state at large $V/t$, 
and the transition point $V_\textrm{cr}/t$ increases 
with increasing $1/K$~\cite{Robaszkiewicz83Physica,Bendt84PRB}. 

When $U/t$ is finite, the spin degree of freedom (in the original electronic model) becomes active. 
The model have been treated by different numerical techniques~\cite{Ung94PRL,Riera00PRB,Clay03PRB,Kuwabara03JPSJ} 
as well as by the bosonization method~\cite{Kuwabara03JPSJ,Sugiura03JPSJ}. 
As discussed in the previous subsections, 
the latter treatment can clarify the physical origin of each state which is realized. 
A phase Hamiltonian is derived following the treatment in $\S$~\ref{subsec_boson}, 
where nonlinear terms additional to ${\cal H}_{\rho}$ in eq.~(\ref{eqn:boson}) and ${\cal H}_{\sigma}$
are generated in the presence of the lattice dimerization $u_\textrm{d}$ and tetramization $u_\textrm{t}$, 
with a cost of lattice elastic energy ${\cal H}_\textrm{el}$. 
These can be expressed as 
\begin{align} 
 {\cal H}_\textrm{d} &= - \frac{g_{1/2} u_\textrm{d} }{2 (\pi \alpha)^2} 
\int \d x \sin 2 \theta_\rho,~\label{eqn:dimerization} \\ 
 {\cal H}_\textrm{t} &= - \frac{g_\textrm{t} u_\textrm{t} }{2 (\pi \alpha)^2} 
\int \d x \sin ( \theta_\rho - \chi_\textrm{t} ) \cos \theta_\sigma,~\label{eqn:tetramization}
\end{align}
when the lattice distortion is parametrized as 
$u_i= u_\textrm{d} \cos{(\pi x_i/a)}+ u_\textrm{t} \cos{(\pi x_i /2a  + \chi_\textrm{t})}$
where $\chi_t$ is a phase factor. 
The term in eq.~(\ref{eqn:dimerization}) is the $4\kf$-Umplapp term 
having the form identical to eq.~(\ref{eqn:4kfUmklapp}) except the presence of $u_\textrm{d}$. 
Therefore provided $u_\textrm{d} \neq 0$ 
it can actually produce the dimer Mott insulator. 
On the other hand, in eq.~(\ref{eqn:tetramization}), 
$g_\textrm{t} \propto t$
couples the two phase variables for charge ($\theta_\rho$) and spin ($\theta_\sigma$). 
In the weakly correlated regime 
this term shows the instability of the system toward the 2$\kf$ CDW state~\cite{FukuyamaTakayama}.

In the strongly correlated regime, on the other hand, the competition between 
the CO insulator ($\theta_\rho=0$ or $\pi/2$) and the dimer Mott insulator ($\theta_\rho=\pi/4$) 
arises due to the $8\kf$-Umklapp term in $\cal{H}_\rho$ and the $4\kf$-Umklapp term 
eq.~(\ref{eqn:dimerization}), respectively. 
This is similar to the case in $\S$~\ref{subsec_boson} 
for ${\cal H}_{\rm 1D}$  in eq.~(\ref{eqn:extHub1D}) with $\delta_\textrm{d} \neq 0$, 
although here, $u_\textrm{d}$, and consequently the presence of the $4\kf$-Umklapp term itself,
should be determined self-consistently.   
The spin degree of freedom in these two states 
is derived to be described by the same phase Hamiltonian 
${\cal H}_{\sigma}'={\cal H}_{\sigma} 
- g_\textrm{t} u_\textrm{t} \int {\rm d}x \cos{\theta_\sigma} 
+ {\cal H}_\textrm{el}$, 
which is identical to that of the SP model~\cite{Fukuyama86Book}. 
Therefore the two insulating states above are 
both unstable toward the non-magnetic SP state with lattice tetramization, 
$u_{\rm t} \neq 0$, 
resulting in coexistent states. 

\begin{figure} 
\centerline{\includegraphics[width=6truecm]{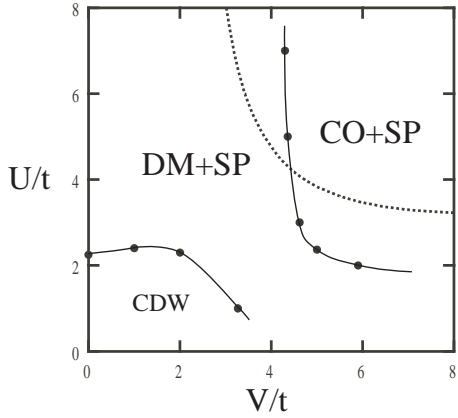}}
\vspace*{-1em}
\caption{Ground state phase diagram of the one-dimensional extended Peierls-Hubbard model 
at quarter-filling 
calculated by the density matrix renormalization group method~\cite{Kuwabara03JPSJ},  
on the plane of $U/t$ and $V/t$ for fixed values of $1/K=0$ 
(dotted line, equivalent to Fig.~\ref{fig_1dEHM}) and $1/K=1$ (filled line).
CO-SP, DM-SP, and CDW phases 
denote the coexistent state of charge order and spin-Peierls lattice tetramization, 
that of dimer Mott insulator and spin-Peierls lattice tetramization, 
and the weak coupling Peierls-Fr\"{o}lich state.
[After ref.~\ref{Kuwabara03JPSJ}.]
}
\vspace*{-1.5em}
\label{fig_Kuwabara}
\end{figure}

These symmetry broken states are in fact reproduced in the numerical calculations. 
Let us show a ground state phase diagram by Kuwabara \et~\cite{Kuwabara03JPSJ} 
determined by the numerical DMRG method in Fig.~\ref{fig_Kuwabara}. 
The spatial modulation of the lattice distortion at each site is determined self-consistently 
and system sizes up to $L=36$ are needed to diminish the finite-size effect. 
One can see that all the states predicted in the phase Hamiltonian 
appear in different regions of parameters $(U/t,V/t)$. 
For the actual pattern of charge density and lattice distortion in each state, 
see ref.~\ref{Kuwabara03JPSJ}. 

There is another type of coupling to the lattice degree of freedom 
from the Peierls coupling above: the Holstein coupling, 
which describes modulation of the one-particle on-site energy 
at the cost of elastic energy. 
The origin of this coupling can be the electron-molecular vibration (e-mv) coupling 
frequently discussed in molecular systems~\cite{NoteEMV}. 
This coupling is usually not large such that the nature of 
phenomena due to strong correlation discussed in this review is qualitatively modified. 
However, recently, exceptionally large deformation of molecules themselves coupled to 
charge disproportionation in (EDO-TTF)$_2$PF$_6$ is observed~\cite{Ota02JMC}, 
suggesting large e-mv coupling in this material~\cite{Drozdova04PRB}.
Another source for the Holstein-type coupling recently considered affecting the CO state 
is the influece of the $X^-$ unit 
implied by experiments in TM$_2X$~\cite{Monceau01PRL,Yu04PRB}.  
The ferroelectricity observed below the CO temperature 
is interpreted as the combination of CO state 
in the dimerized stack of TM molecules (see Fig.~\ref{fig_1dMF2}) 
and the $\boldsymbol{q}=0$ motion of the anions, 
producing large electrical polarization. 

Numerical ED calculations for small cluster sizes up to about $L=16$ sites 
including this Holstein coupling as well as the Peierls coupling suggests that the former type 
works toward stabilization of the CO state in general~\cite{Clay03PRB,Riera01PRB}. 
We note that since the lattice is 3D in nature, 
these couplings would result in finite temperature phase transitions~\cite{Sugiura03JPSJ}, 
which are not fully investigated yet. 
Furthermore, the inclusion of intrinsic dimerization $\delta_\textrm{d}$ 
in the model might modulate the results, 
such as a further stabilization of the dimer Mott insulating state~\cite{Riera01PRB,Sugiura03JPSJ}. 
Still, systematic works which would provide 
unified pictures of the quasi-1D electron-lattice coupled quarter-filled systems,
together with the interchain coupling discussed in $\S$~\ref{subsec_interchain}, are to be done~\cite{SeoInprep2}. 

\subsection{Geometrical frustration}\label{subsec_frustration}

\vspace{3pt}

As introduced in $\S$~\ref{subsec_interchain}, 
the interchain Coulomb interaction $V_\perp$ added to the 1D EHM 
stabilizes the 2D or the 3D CO state with an anti-phase pattern between chains. 
However, in the actual molecular materials  
the manner of coupling between chains is frequently more anisotropic. 
For example, in TM$_2X$, as shown in Fig.~\ref{fig_DCandTM}, 
three kinds of interchain bonds are formed in a zigzag way. 
Then the interchain Coulomb repulsion $V_{b}$ favors the CO pattern 
along the $b$-direction to be anti-phase between chains,  
while $V_{p1}$ and $V_{p2}$ favor the in-phase pattern. 
(The observed ferroelectricity in ref.~\ref{Monceau01PRL} points to the latter.) 
Since these Coulomb energies are expected to have similar values~\cite{Fritsch91JPIF,Castet96JPIF}, 
the two CO patterns compete with each other. 
This is an example of geometrical frustration. 

Effects of the geometrical frustration are intensively studied in 
localized spin systems~\cite{Ramirez01Book} defined on lattice structures 
with odd number of sites when rounding along the bonds of a cell, e.g., triangular forms 
as in the above example of TM$_2X$. 
It is known to drive the system toward destabilization of the classical antiferromagnetic state
and sometimes even result in a total destruction of it, namely, in a spin disordered state. 
The analogy between the classical picture of CO and the spin systems  
noted in $\S$~\ref{subsec_1dEHM} 
naturally leads us to the problem of geometrical frustration in CO systems, 
where the CO may be destabilized as well. 
This was first considered by Anderson~\cite{Anderson56PR} 
in the context of the ``classical" 3D CO system, 
magnetite Fe$_3$O$_4$, described as an Ising model 
on the phyloclore lattice which is a typical frustrated lattice structure. 
%
\begin{figure}
\centerline{\includegraphics[width=5true cm]{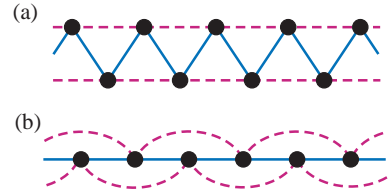}}
\vspace*{-1em}
\caption{(Color online) 
(a) Zigzag ladder and (b) one-dimensional chain with next nearest neighbor 
interactions. These two are equivalent with each other as seen in the figure. 
}
\vspace*{-1.5em}
\label{fig_zigzag}
\end{figure}

Treating the zigzag coupling in the 2D model for TM$_2X$ with full anisotropy is a problem yet to be solved. 
A similar situation will be discussed in $\S$~\ref{subsec_ATL}, 
related to the CO in quasi-2D $\theta$-ET$_2X$. 
Instead, here we consider a more simple system but having many common aspects: 
a double-chain system with zigzag couplings between them, 
i.e., a zigzag ladder, shown in Fig.~\ref{fig_zigzag}. 
As seen in Fig.~\ref{fig_zigzag}, it is equivalent to a 1D model with next-nearest-neighbor 
interactions. 
Then the EHM on this chain structure can be written as 
${\cal H}_{\rm 1D}$ in eq.~(\ref{eqn:extHub1D}) with $\delta_{\rm d}=0$ 
added by next-nearest terms, 
$t_2 \sum_{i\sigma} ( c^{\dagger}_{i\sigma}c_{i+2\sigma} + {\rm h.c.} ) 
+ V_2 \sum_i n_in_{i+2}$. 
In the following we write the nearest neighbor terms as $t_1$ and $V_1$. 

This model at quarter-filling has been studied as an effective model 
for two different material systems. 
One is for TMTSF$_2X$, with $t_2=0$, 
extending the range of Coulomb interaction terms from the previous works~\cite{Kobayashi98JPSJ,Yoshioka01JPSJ}. 
Another is for the Cu-O subunit of a transition metal oxide PrBa$_2$Cu$_4$O$_8$, 
where the effective model for the Cu sites has a zigzag ladder structure 
with $t_1 \ll t_2$, and $V_1 \simeq \sqrt{2} V_2$ suggested by the Cu-Cu distances~\cite{Seo01PRB,Nishimoto03PRB}. 
For $t_1=0$ or $t_2=0$, the model at $U=\infty$ is exactly mapped onto XXZ models 
when one follows the mapping of Ovchinnikov explained in $\S$~\ref{subsec_1dEHM},  
therefore the analogy between the frustrated localized spin system 
and the CO system is straightforward here again. 

For the $t_2=0$ case, 
Emery and Noguera~\cite{Emery88PRL} actually studied the 1D XXZ model with next nearest $J_z$ term, 
while Yoshioka \et~\cite{Yoshioka01JPSJ} 
studied the finite-$U$ case by the method described in $\S$~\ref{subsec_boson}. 
In these studies the interaction terms are treated perpurbatively from the weak-coupling regime. 
They have discussed the competition between two different CO patterns 
favored respectively by $V_1$ and $V_2$. 
The one favored by $V_2$ having wave vector 2$k_\textrm{F}$ ($\kf$ defined for $t_2=0$) 
shows 2$\kf$ SDW correlation developped. 
The co-existent state of 2$k_\textrm{F}$ SDW and 2$k_\textrm{F}$ CDW 
was first proposed based on the MF approximation~\cite{Kobayashi98JPSJ}, 
to discuss such state observed in TMTSF$_2X$~\cite{Ravy04CR}. 
We note that this ``2$k_\textrm{F}$" CDW is 
different from the Peierls-Fr\"{o}lich state due to electron-lattice coupling, 
but it is originated from the strong correlation, $V_2$, i.e., it is indeed a CO state. 
A possible insulating bond ordered wave phase 
is also found in the above weak-coupling approaches, 
in the competing region.
\begin{figure}
\centerline{\includegraphics[width=7.0true cm]{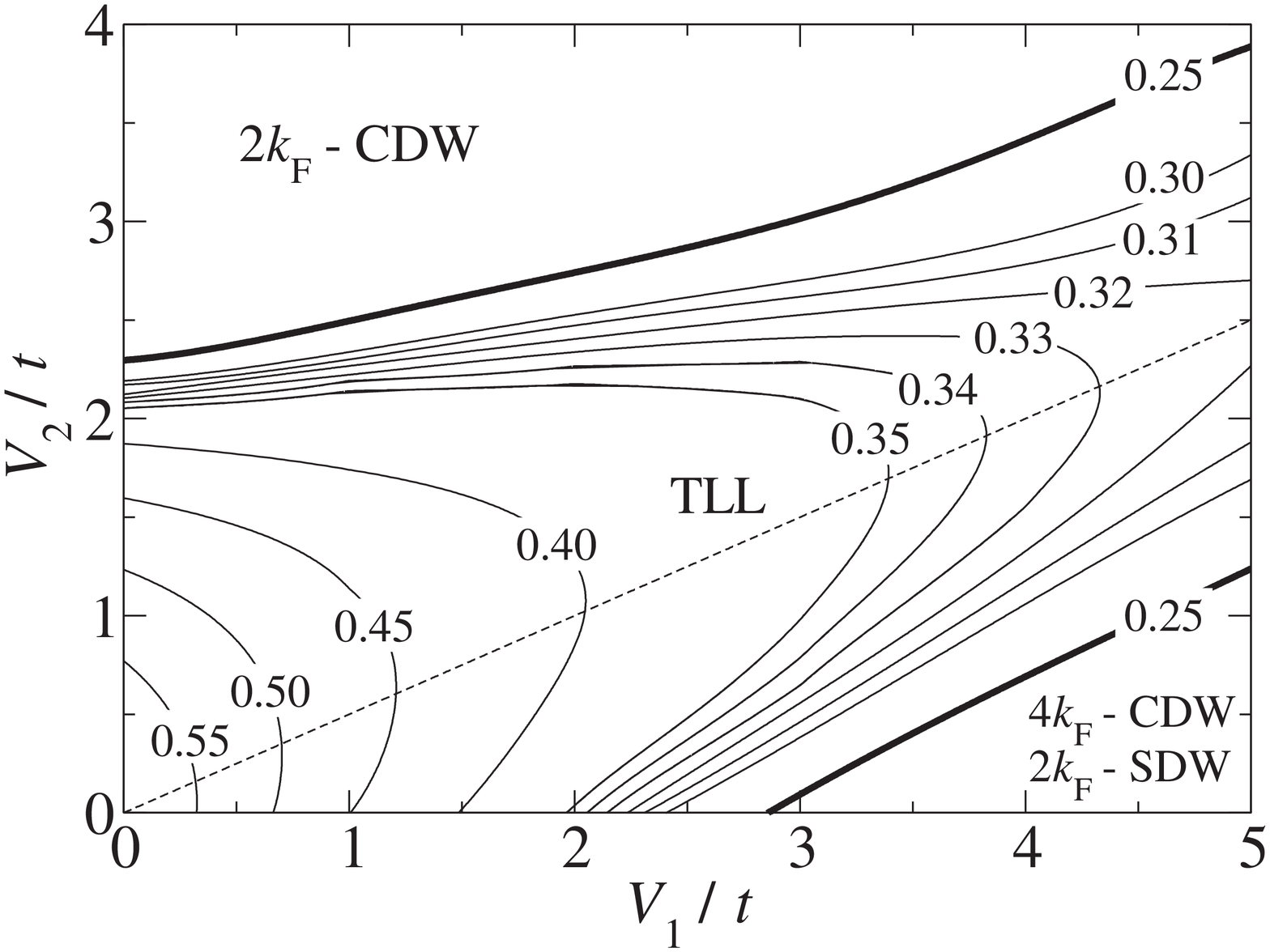}}
\vspace*{-1em}
\caption{Ground state phase diagram of the extended Hubbard model 
at quarter-filling on the zigzag ladder, determined by 
the density matrix renormalization method.~\cite{Ejima05PRB} 
$U/t_1$ is fixed at 10 while $t_2$ is set to 0. 
Sandwitched by the two insulating CO phases with different charge patterns, 
the Tomonaga-Luttinger liquid (TLL) metallic phase is extended 
toward the frustrated line $V_1 = 2V_2$. 
The contour lines indicate the TLL parameter $K_\rho$.  
[Reprinted figure with permission from 
S. Ejima {\it et al.}, Phys. Rev. B {\bf 72} (2005) 033101.   
Copyright (2005) by the American Physical Society.]
}
\vspace*{-1.5em}
\label{fig_zigzag_phased}
\end{figure}

Numerical studies on such models on the zigzag ladder have found 
a different aspect at large $V_{ij}$'s where the weak coupling approaches 
above may break down: 
a wide region of the metallic TLL phase between the two CO phases~\cite{Seo01PRB}. 
A phase diagram by the numerical DMRG method is shown in Fig.~\ref{fig_zigzag_phased}~\cite{Ejima05PRB}, 
where TLL phase is widely realized in the $V_1 \simeq 2V_2$ region, 
with $K_\rho$ close to 1/4 suggesting the closeness to CO. 
This is a consequence of the geometrical frustration, 
and in such metallic phase with both $V_1$ and $V_2$ having large values, 
the existence of large fluctuations of both CO states is expected~\cite{SeoInprep}. 

Recently, it is pointed out that 
the CO state in DI-DCNQI$_2$Ag is affected by similar geometrical frustration. 
This is due to a peculiar situation where  
the interchain configuration along each plaquette in the $a$-$b$ plane 
(see Fig.~\ref{fig_DCandTM}) 
has a spiral symmetry which is not compatible with the twofold periodicity 
generated by CO~\cite{Kanoda05JP4}. 
Such frustration due to the spiral symmetry 
is also found to exist in the process of 
the metal-insulator transition in DMe-DCNQI$_2$Cu~\cite{Tazaki}, 
where the $\pi$-electron couples to the Cu $d$-electron 
producing a three-fold periodic structure~\cite{RKato}
again not compatible with the spiral symmetry, 
which is beyond the scope of this review. 

\vspace{3pt}

\section{Quasi-Two-Dimensional Systems}\label{sec_q2d}

\vspace{3pt}

Many theoretical works on CO states in 2D models near half-filling 
have been performed, such as in the Hubbard model and in the $t$-$J$ model 
on the square lattice, 
mainly devoted to the high-$T_{\rm c}$ SC cuprates 
and related transition metal oxides~\cite{Carlson04Book}. 
However, studies on the quarter-filled case are mostly 
triggered by the recent experimental findings of CO in quasi-2D molecular conductors. 
To the authors' knowledge, only a few have been done beforehand.  
Ohta \et~\cite{Ohta94PRB} studied the 2D EHM on the square lattice 
for different fillings including one quarter,  
in fact motivated by the cuprates, 
while a more specific case was studied~\cite{Seo98JPSJ}  
to explain the CO transition in NaV$_2$O$_5$ 
(see $\S$~\ref{sec_related}). 

Since CO has been found in many quasi-2D $A_2B$ systems such as in ET$_2X$, 
theoretical works have been performed on the 2D 
EHM adopting the full anisotropy (or a part of it) of the materials. 
On the other hand, the quarter-filled EHM on the square lattice 
has been studied as well from a rather general viewpoint. 
Such 2D models are difficult to analyze theoretically in a controlled way, 
and intensive efforts are now in progress. 
We will review the present status in this section. 

\subsection{Charge ordering in quasi-two-dimensional systems}\label{subsec_COq2d}

\vspace{3pt}

\begin{figure}
\centerline{\includegraphics[width=7.0truecm]{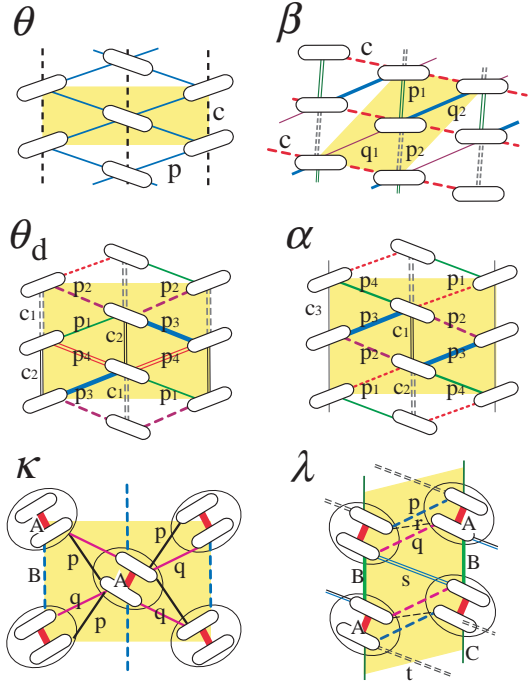}}
\vspace*{-1em}
\caption{(Color online) 
Schematic representation of the 2D layers in 
different polytypes of quasi-2D $A_2B$ materials. 
The indices are not only for $t_{ij}$ but also for different values of $V_{ij}$, 
although the degree of anisotropy can be different (see text). 
In the $\kappa$-type and the $\lambda$-type structures 
the thick $A$ bonds are those with the transfer integral $t_A$ 
considerably larger than the others, i.e., dimers. 
[By courtesy of C. Hotta.]
}
\vspace*{-1.5em}
\label{fig_polytypes}
\end{figure}

Direct observation of the CO transition in quasi-2D $A_2B$ compounds 
was first in $\theta$-ET$_2$RbZn(SCN)$_4$~\cite{Miyagawa00PRB,Chiba01JPCS}, 
and next in $\alpha$-ET$_2$I$_3$~\cite{Takano01JPCS}, 
both found in $^{13}$C-NMR experiments. 
Now many quasi-2D $A_2B$ materials, including non-ET based compounds, 
are recognized to show CO.~\cite{TakahashiThisVol}
Their crystal structures are composed of alternating layers of $A$ molecules and $B$ units, 
and quasi-2D electronic states are realized; 
$t_{ij}$ along the interlayer direction are small due to the insulating $B$ layers. 
These materials have a variety in their 2D arrangements of $A$ molecules, 
sometimes even for the same chemical formula, i.e., polytypes, 
classified by greek characters, 
$\alpha$, $\beta$, $\kappa$, $\theta$, $\lambda$, and so on~\cite{TMori98BCSJ}. 
Some representative examples are shown in Fig.~\ref{fig_polytypes}, 
for which the actual values of transfer integrals calculated by the extended H\"{u}ckel method 
we refer to, e.g.  refs.~\ref{Seo04CR} and \ref{TMori98BCSJ}. 
Because of such anisotropy in the lattice structures, 
the EHM we should treat, ${\cal H}_{\rm EHM}$ in eq.~(\ref{eqn:extHub}), would be complicated. 
As was mentioned in $\S$~\ref{sec_Intro}, 
we emphasize that such diversity of lattice structures here 
is one of the important characteristics producing the rich variety of properties 
in molecular conductors. 
CO is in fact very much affected by such geometry of the lattice as we will see in the following subsections. 

A theoretical approach to handle this diversity was pursued by Kino and Fukuyama~\cite{Kino96JPSJ}, 
who considered Hubbard models ($V_{ij}=0$) for ET$_2X$, 
taking into account the full anisotropy in $t_{ij}$ for each polytype,  
and applied MF approximation to the on-site Coulomb interaction term $U$. 
By this 
they have provided a way to search for relations between 
the crystal structure and electronic properties, 
and also for generic understandings even among different polytypes~\cite{Kino95JPSJ_3}. 
Since the polytypes show a variety, different ways of such systematic views  
have been discussed by different authors since then~\cite{Seo04CR,TMori98BCSJ,Hotta03JPSJ}. 
One way is to consider an anisotropic triangular lattice~\cite{Kino95JPSJ_2,McKenzie97Science}, 
refined by Hotta~\cite{Hotta03JPSJ} as follows. 

In Fig.~\ref{fig_polytypes}, one can see that $\theta$, $\alpha$, and $\beta$-type 
structures have the triangular lattice as basis, ``colored" with anisotropy. 
Since the charge transfer is realized as $[$ET$_2]^{+}X^{-}$, 
the one-particle HOMO bands as a whole are quarter-filled in terms of holes. 
Effective models for the materials with these symmetries would be the quarter-filled 
EHM on an anisotropic triangular lattice, 
with $V_{ij}$ having the same anisotropy as $t_{ij}$. 
For example, for the $\theta$-type structure with highest symmetry among the polytypes, 
\begin{align}
{\cal H}_\theta = &-t_p \sum_{\langle ij \rangle_p,\sigma} (c^\dagger_{i \sigma} c_{j \sigma} +h.c.) 
-t_c \sum_{\langle ij \rangle_c,\sigma} (c^\dagger_{i \sigma} c_{j \sigma} +h.c.) \nonumber\\ 
&+ U \sum_{i} n_{i\uparrow}n_{i\downarrow}
+ V_p \sum_{\langle ij \rangle_p} n_i n_j + V_c \sum_{\langle ij \rangle_c} n_i n_j, 
\label{eqn:theta}
\end{align}
where $\langle ij \rangle_p$ and $\langle ij \rangle_c$ denotes the 
site pairs $i$ and $j$ along the $p$ bonds and the $c$ bonds, respectively. 
In contrast, $\kappa$ and $\lambda$-type stuctures 
consist of dimers, 
i.e., pairs of molecules connected by $t_{ij}$ considerably larger than the others, 
indicated as the $A$ bonds in Fig.~\ref{fig_polytypes}; 
if one considers the dimers as units, 
these structures are topologically equivalent to $\theta$ and $\beta$-type structures, respectively. 
Half-filled models on the anisotropic triangular lattice can be effective models 
for such systems with large dimerization. 
Then the on-dimer Coulomb repulsion $U_\textrm{dimer}$ can lead the system toward 
the dimer Mott insulator (see refs.~\ref{Seo04CR}, \ref{Kino96JPSJ}-\ref{Miyagawa04CR}). 

The CO states are stabilized in the former quarter-filled models. 
In Kino and Fukuyama's work on the Hubbard model for $\alpha$-ET$_2$I$_3$~\cite{Kino95JPSJ}, 
an insulating state accompanied with charge disproportionation between stacks 
(the vertical stripe type CO pattern; see $\S$~\ref{subsec_ATL}) is found. 
However, the charge pattern there is not consistent with later experiments. 
Moreover in a MF study on the Hubbard model for the other CO material $\theta$-ET$_2$RbZn(SCN)$_4$,
CO state is not even stabilized~\cite{Seo98JPSJ_2}. 
Following these works, 
$V_{ij}$ added to the Hubbard model retaining the full anisotropy in $t_{ij}$, 
i.e., the EHM for ET$_2X$ on the anisotropic triangular lattice, 
has been introduced by Seo~\cite{Seo00JPSJ}, 
who discussed the CO states there based on the MF approximation. 
As in the quasi-1D systems discussed in $\S$~\ref{sec_q1d}, 
the intersite Coulomb interaction is also crucial for CO in the quasi-2D materials. 

The values of $V_{ij}$ are in fact estimated to be appreciable compared to $U$. 
As in the quasi-1D compounds in $\S$~\ref{sec_q1d}, 
$U$ is believed to be of the order of 1 eV 
in the ET compounds as well, 
while, again, the actual values for $V_{ij}/U$ estimated as $0.2 \sim 0.7$~\cite{Ducasse97SM,Imamura99JCP,TMori00BCSJ} 
are too ambiguous for quantitative arguments. 
Typical values for $|t_{ij}|$ in the ET materials 
estimated by the extended H\"{u}ckel method 
range around $0.1 \sim 0.25$ eV~\cite{Seo04CR,TMori98BCSJ}. 
Therefore these materials are indeed strongly correlated systems.  
We note again that the degree of anisotropy in $t_{ij}$ 
and that in $V_{ij}$ does not correspond in a one-to-one manner. 
For example in many members of $\theta$-ET$_2X$ described by eq.~(\ref{eqn:theta}), 
$|t_c|$ is much smaller than $|t_p|$~\cite{HMori98PRB}, 
despite of their close inter-molecular distance
resulting in $V_c \simeq V_p$~\cite{TMori00BCSJ}. 

The anisotropic triangular lattice is somewhat complicated 
and therefore it is not only tough to handle it theoretically 
but also sometimes difficult to extract explicit key parameters for the physics therein. 
One natural way of thinking, as was pointed out by McKenzie \et~\cite{McKenzie01PRB}, 
is to study the quarter-filled EHM on the square lattice, 
which is much simpler but would still show essential features 
of the CO transition in 2D, 
and even of the expected SC state~\cite{Merino01PRL} 
($\S$~\ref{subsec_SC}). 
In this model, there is one kind of transfer integral $t$ 
and also inter-site Coulomb repulsion $V$, 
along nearest-neighbor sites both in $x$ and $y$ directions. 
Such studies on the square lattice can shed light on 
the properties of the anisotropic triangular lattice EHM, 
as the former is one of the limiting cases of the latter. 
For example, in the $\theta$-type structure 
if we set $t_c=0$, $V_c=0$ and $t=t_p$, $V=V_p$, 
${\cal H}_\theta$ in eq.~(\ref{eqn:theta}) becomes equivalent to the square lattice EHM. 
Neglecting $t_c$ is in fact a good approximation for $\theta$-ET$_2X$ 
whereas $V_c$ cannot be neglected, as mentioned above, 
while this square lattice case might be applied to some $\beta''$-type compounds~\cite{Merino01PRL}. 
Note that electron-hole symmetry holds in the square lattice EHM 
while $t_c$ breaks it, so relative signs between $t_{ij}$ become distinct. 
We will explain studies on the square lattice EHM first ($\S$~\ref{subsec_SqL}) and 
then the anisotropic triangular lattice case next ($\S$~\ref{subsec_ATL}), 
although the research has been developped more or less in a parallel way.  

\subsection{Extended Hubbard model on square lattice}\label{subsec_SqL}

\vspace{3pt}

\begin{figure}
\centerline{\includegraphics[width=5.0truecm]{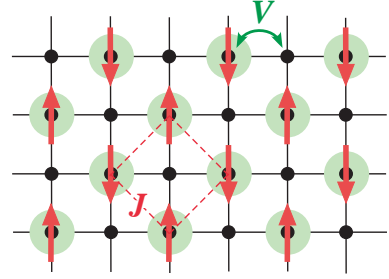}}
\caption{(Color online) 
Checkerboard charge ordered insulating
ground state of the square lattice extended Hubbard model at quarter filling 
in large $U/t$ and $V/t$. 
An antiferromagnetic interaction $J$ occurs between spins along the diagonals
shown in the figure. 
}
\vspace*{-1.5em}
\label{fig_CO2D}
\end{figure}

The existence of a checkerboard type CO 
with wave vector ${\boldsymbol q}$=$(\pi/a,\pi/a)$
on the quarter-filled square lattice EHM, as shown in Fig.~\ref{fig_CO2D}, 
has been first demonstrated by Ohta \et~\cite{Ohta94PRB}, 
for $U=8t$ and $V=3t$. 
It is based on calculations of 
the equal-time charge correlation function 
$C({\boldsymbol q})=L^{-1} \sum_{ij} \langle n_i n_j \rangle {\rm e}^{i {\boldsymbol q} \cdot {\boldsymbol R}_{ij}}$, 
where $L$ is the total number of sites in the square shaped clusters 
and ${\boldsymbol R}_{ij}$ is the vector connecting site $i$ and $j$, 
by use of the Lanczos ED technique with cluster sizes up to $L=16$. 
This pattern can be understood naturally  
as the extention of the 1D case (Fig.~\ref{fig_COvsDM}(a))
from the idea of ``Wigner crystal on lattice". 

They have also derived an effective spin model in such CO state 
by the fourth order perturbative expansion from strong coupling, $t \ll (U,V)$. 
It is an antiferromagnetic Heisenberg model on the square lattice, 
rotated 45 degrees from the original lattice (see Fig.~\ref{fig_CO2D}), 
with a nearest neighbor exchange coupling, 
\begin{align}
J = {4 t^4 \over 9 V^2} \left( \frac{4}{U} + \frac{1}{V} + \frac{4}{U+4V} \right). \label{eqn:Jfourth}
\end{align}
Note that even at $U/t=\infty$, 
novel ring exchange processes break the zeroth order spin degeneracy 
resulting in a finite $J=4t^4/9V^3$.~\cite{McKenzie01PRB} 
Since the next-nearest-neighbor coupling $J'$ is estimated to be much smaller, 
an antiferromagnetic spin order as shown in Fig.~\ref{fig_CO2D} is expected at the ground state 
since the square lattice Heisenberg model shows N\'{e}el order~\cite{Manousakis91RMP}.

\begin{figure}
\centerline{\includegraphics[width=7.0truecm]{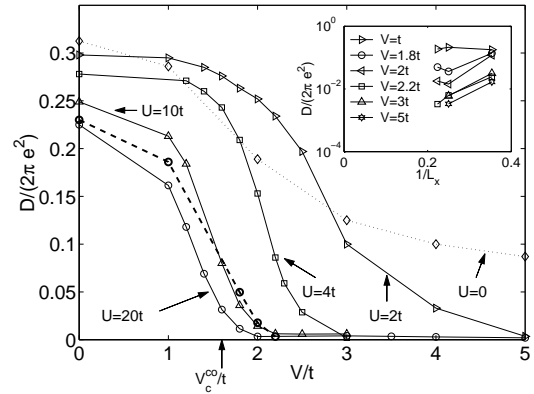}}
\vspace*{-1em}
\caption{
The Drude weight, $D$, for the quarter-filled square lattice extended Hubbard model 
as a function of  $V/t$, for $L=16$ and various values of $U/t$
(continous and dotted lines), and for $L=20$ and $U=10t$ (dashed line)~\cite{Calandra02PRB}.
The arrow in the horizontal axis marks the
onset of checkerboard charge ordering for $U=10t$. 
The inset shows 
the finite-size scaling of $D$ as a function
of $1/L_x$ for different values of $V/t$ with $U=10t$.
The metal-insulator transition occurs at $V_{\rm cr}^{\rm MI} \simeq 2.2t$. 
[Reprinted figure with permission from 
M. Calandra {\it et al.}, Phys. Rev. B {\bf 66} (2002) 195102. 
Copyright (2002) by the American Physical Society.]
}
\vspace*{-1.5em}
\label{fig:drude}
\end{figure}

A systematic numerical Lanczos ED study was demonstrated by Calandra \et~\cite{Calandra02PRB}, 
for cluster sizes up to $L=20$. 
In this case, a powerful method to determine from small systems 
whether the bulk system is metallic or insulating 
is to evaluate the Drude weight $D$~\cite{Kohn64PR,Stephan90PRB,Scalapino93PRB}.
It is given by 
\begin{equation}
\frac{D}{2 \pi e^2} = - \frac {\langle 0|K|0\rangle}{4 L}  - \frac{1}
{L} \sum_{n \ne 0} \frac{|\langle n|j_x|0\rangle|^2}{E_n-E_0}, 
\label{eqn:drude}
\end{equation}
where $E_0$ and $E_n$ denote the ground state ($|0\rangle$) and excited state  ($|n\rangle$) energies
of the system, respectively.~\cite{Dagotto94RMP} $K$ is the kinetic energy operator
and $j_x$ is the current operator in the $x$ direction
at zero wavevector (${\boldsymbol q}=0$). 
The occurrence of an insulating phase is marked by the
exponential vanishing of $D$ with the linear size of the system
$L_x = \sqrt{L}$ \cite{Kohn64PR,Scalapino93PRB}.
In Fig. \ref{fig:drude} the results are shown, 
where $D$ is plotted as a function of $V/t$ for different $U/t$. 
As $V/t$ is increased, $D$ decreases until it eventually vanishes. 
For $U/t=10$ 
the critical value for the metal-insulator transition 
is estimated as $V_{\rm cr}^\textrm{MI}/t \simeq 2.2$. 
We should note that this might not be conclusive as an accurate value 
as the finite size scaling shown in the inset of Fig. \ref{fig:drude} 
is only for the three clusters $L$=8, 16, and 20, due to computorwise limit. 
Nevertheless, for $V > V_{\rm cr}^\textrm{MI}$, 
$D$ displays an exponential dependence with $1/L_x$ 
as expected for an insulator\cite{Kohn64PR,Scalapino93PRB}, 
while for $V < V_{\rm cr}^\textrm{MI}$, in contrast, 
it weakly depends on $1/L_x$
extrapolated to a finite value in the thermodynamic limit,
consistent with a metallic state.
On the other hand, the occurrence of CO has been investigated by computing
$C(\boldsymbol{q})$ systematically. 
The extrapolation suggests $V_{\rm cr}^\textrm{CO}/t \simeq 1.6$, 
where $C(\pi/a,\pi/a)$ starts to show a peak as $V$ is increased, 
which possibly overestimate its value due to finite size effects. 
When $U$ decreases the critical value $V_{\rm cr}^\textrm{CO}$ increases, 
similarly as in the 1D case (see Fig.~\ref{fig_1dEHM}). 

Since the estimates above provide $V_{\rm cr}^\textrm{MI} > V_{\rm cr}^\textrm{CO}$, 
a possible CO metallic phase, not found in the 1D EHM, is suggested.  
In fact, a recent 
variational Monte-Carlo calculation~\cite{WatanabeInprep} (see next subsection) 
have confirmed its existence, 
with a CO metallic phase realized between the uniform metallic and the CO insulating phases, 
i.e., at $V_{\rm cr}^\textrm{CO}< V < V_{\rm cr}^\textrm{MI}$.  
We note that another work using projector correlated method~\cite{Hanasaki05JPSJ} 
shows that such feature is also seen at finite temperatures. 
This is in clear contrast with the purely 1D EHM discussed in 
$\S$~\ref{sec_q1d}, 
where the quantum fluctuation destroys the finite-$T$ CO phase transition. 
More theoretical efforts are needed, 
to explore the overall phase diagram for this model, 
in the $(U,V,T)$ parameter space, 
for the critical values of CO, metal-insulator, and antiferromagnetic 
phase transitions.

Although the analysis above is restricted to numerics, 
the $U/t=\infty$ limit has been investigated analytically 
by McKenzie \et~\cite{McKenzie01PRB}. 
In this limit the square lattice EHM reads after projecting out the doubly occupied sites as
\begin{equation}
 {\cal H}_{tV}=-t\sum_{\langle ij \rangle\sigma} {\cal P} \left(c^{\dagger}_{i\sigma}
    c_{j\sigma}+h.c.\right) {\cal P} 
    +V \sum_{\langle ij \rangle} n_in_j,
\label{eqn:extHubp}
\end{equation} 
where ${\cal P}$ projects out the doubly occupied configurations 
due to their prohibition at $U/t=\infty$.  
We refer to this model as the $t$-$V$ model; 
this is the generalization of the $U/t=\infty$ limit of the 1D case 
discussed in $\S$~\ref{subsec_1dEHM}, 
whereas the charge and spin degree of freedoms cannot be decoupled in 2D. 

They have introduced an SU($N$) generalization of this model 
in which the spin index runs from 1 to $N$, and carried out a slave boson theory, 
which have been applied to various models for strongly correlated 
electron systems by different authors~\cite{Newns87Review}. 
The electron creation operator is replaced by 
$c^{\dagger}_{i\sigma}=f^{\dagger}_{i \sigma} b_i$, where the
spinless charged boson operator $b_i$ is introduced to keep
track of the empty sites, and $f^{\dagger}_{i \sigma}$ is a fermion
operator carrying spin. In order to preserve
the anticommutation relation for the electrons
the new operators must satisfy the local constraint
$f^{\dagger}_{i\sigma}  f_{i\sigma}+ b^{\dagger}_i b_i=N/2$. 
Hereafter, whenever a repeated 
$\sigma$ index appears a sum from 1 to $N$ is assumed. 
Note that the original EHM has $N=2$. 

In the coherent state path integral formulation~\cite{Kotliar88PRL} 
the Lagrangian at imaginary time $\tau$ is given by
\begin{align} 
L(\tau) &= \sum_{i} \left( f^{\dagger}_{i\sigma}( \partial_\tau - \mu) f_{i\sigma} 
+ b^{\dagger}_i\partial_\tau b_i \right) \nonumber \\
- {t \over N} &\sum_{\langle ij \rangle} 
(f^{\dagger}_{i\sigma} f_{j \sigma} b^{\dagger}_j b_i + h.c. )
+{V \over N} \sum_{\langle ij \rangle} 
f^{\dagger}_{i\sigma} f_{i\sigma} f^{\dagger}_{j\sigma'}
f_{j\sigma'} \nonumber \\
&+ \sum_{i} i \lambda_i (f^{\dagger}_{i\sigma}f_{i\sigma} +
 b^{\dagger}_i b_i - N/2),
\label{lagran}
\end{align}
where $\mu$ is the chemical potential 
and $\lambda_i$ is a static Lagrange multiplier enforcing the local constraint above.
We have used the fact that 
$c^{\dagger}_{i\sigma}  c_{i\sigma}=f^{\dagger}_{i\sigma}  f_{i\sigma}$. 

Following Kotiar and Liu's work on the square lattice Hubbard model~\cite{Kotliar88PRL}, 
it is convenient to choose the radial gauge to avoid possible infrared divergences 
where the boson amplitude becomes a real number, $r_i=|b_i|$, 
and $\lambda_i$ becomes a dynamical bosonic field: $\lambda_i(\tau)$.
By using the relation $f^{\dagger}_{i\sigma} f_{i\sigma} =
N/2 - b^{\dagger}_i b_i $ to replace one
pair of fermion operators in the $V$ term 
and integrating out the other fermionic degree of freedoms, 
we are left with an effective Lagrangian which is quadratic in the boson fields. 

The MF solution of the effective bosonic model is obtained 
by assuming that the boson fields are spatially homogeneous and time 
independent: $r_i(\tau)=b$ and $i \lambda_i(\tau)=\lambda $. 
This treatment is exact in the $N \rightarrow \infty$ limit, 
therefore is a base for 
the expansion in powers of $1/N$~\cite{Newns87Review,Kotliar88PRL}. 
The resulting MF free energy is 
\begin{equation}
F^\textrm{MF}(b,\lambda)=
-{N \over \beta}  \sum_{{\boldsymbol k}, \omega_n}
 \ln( \epsilon_{\boldsymbol  k} - i\omega_n)
 + \lambda (b^2-{N \over 2} ), 
 \label{MF}
\end{equation}
where $\beta=1/(k_\textrm{B} T)$ and $\omega_n$ is the fermion Matsubara frequency. The
MF eigenenergy is given by 
$\epsilon_{\boldsymbol  k} = - 2 t b^2 ( \cos k_x + \cos k_y ) / N  
+ \lambda - \mu + 4 V n/ N$. 
Minimization of $F^\textrm{MF}$ with respect to
$b$ and $\lambda$ leads to 
$b^2=N/2-n, \lambda = 2t \sum_{\boldsymbol  k} f(\epsilon_{\boldsymbol  k}) ( \cos k_x + \cos k_y + 4 V)$. 
and $\mu$ is adjusted to give the correct electron filling.  
This form, eq.~(\ref{MF}), 
indicates that the MF solution for the large-$N$ generalized $t$-$V$ model 
describes just a renormalized Fermi liquid, 
analogous to the case of the Hubbard model~\cite{Kotliar88PRL}. 
In the case of quarter-filling and $N=2$ 
(which is somewhat artificial as one implicitly consider $N$ to be large), 
the bandwidth is reduced to half its bare value as $b^2 =1/2$ 
and the band is shifted from its bare position by $\lambda$.  
Namely, the mass is enhanced as twice the bare value, i.e., $m^*/m=1/b^2=2$. 
The overall effect of the nearest-neighbour Coulomb interaction, $V$, 
reduces to a constant shift in the chemical potential.

The leading $1/N$ corrections modify the MF solution 
so that when $V/t$ is increased the Fermi liquid phase results 
in an instability toward the checkerboard type CO state. 
This is seen when writing the boson fields in terms of the static MF solution
$(b,\lambda)$ and the dynamic fluctuating parts: $ r_i(\tau)=b + b \delta r_i
(\tau)$, and $i \lambda_i (\tau)= \lambda + i \delta \lambda_i (\tau)$.
The resulting effective action, to the second order in the boson fields, 
is $S = F^\textrm{MF} + S^{(2)}$, where the second term due to the fluctuations
in the boson fields is written in the form as 
\begin{eqnarray}
S^{(2)} = {1 \over 2 \beta } \sum_{{\boldsymbol  q}, \nu_n} \left[
\begin{array}{cc} \delta r(-{\boldsymbol  q}, -\nu_n) &  \delta
\lambda(-{\boldsymbol  q}, -\nu_n) 
\\ \end{array} 
\right] \times 
\nonumber \\
\left( \begin{array}{cc}
\Gamma_{rr} & \Gamma_{r\lambda} \\
\Gamma_{\lambda r} & \Gamma_{\lambda \lambda} \\ \end{array}\right)
\left( \begin{array}{cc}
\delta r({\boldsymbol  q}, \nu_n)  \\
\delta \lambda({\boldsymbol  q}, \nu_n) \\ \end{array}\right),
\label{action}
\end{eqnarray}
where $\nu_n$ is the boson Matsubara frequency; for the explicit form of 
the elements in $\hat{\Gamma}({\boldsymbol  q}, \nu_n)$ see ref.~\ref{McKenzie01PRB}. 
We note that $\Gamma_{\lambda \lambda}$ 
is the Lindhard function describing density-density fluctuations in the renormalized band. 
The propagators of the boson fields 
$\hat{D}({\boldsymbol  q}, \nu_n )= \hat{\Gamma}^{-1} ({\boldsymbol  q}, \nu_n)$ are of the order $O$($1/N$), 
as they should. 
The condition for an instability of the Fermi liquid phase is
when the quadratic form (\ref{action}) becomes negative at some ${\boldsymbol  q}$ for $\nu=0$, 
so that fluctuations in the charge density will decrease the free energy. 
This is determined by $\textrm{det} \hat{\Gamma} < 0$ at ${\boldsymbol  q}=(\pi/a,\pi/a)$ for quarter filling, 
which provides $(V/t)_{\rm cr} = 0.69$ for the critical value from the Fermi liquid to the CO state 
for this large-$N$ $t$-$V$ model. 
We note that the inclusion of diagonal hopping $t'$ 
(equivalent to $t_c$ in eq.~(\ref{eqn:theta})) 
is easily incorporated in this scheme, 
which does not modify the results qualitatively~\cite{McKenzie01PRB}. 

\subsection{Extended Hubbard model on anisotropic triangular lattice}\label{subsec_ATL}

\vspace{3pt}

As we have seen in the previous subsection, 
the intersite Coulomb repulsion 
can drive the system toward CO in 2D systems as well. 
In the strongly correlated regime, 
the spins are described by an effective 2D Heisenberg model 
with spins located on the charge rich sites. 
Such aspects are common with the 1D cases discussed in $\S$ \ref{sec_q1d}, 
where CO results in magnetic properties of 1D Heisenberg chains. 
These knowledges are very useful in treating 
the more anisotropic cases introduced in $\S$~\ref{subsec_COq2d}, 
with full anisotropy taken into account. 
There, calculations beyond MF are still in progress, 
but if we deduce from the MF results together with the picture above, 
qualitative understandings of the experiments can be obtained~\cite{Fukuyama00Physica} as follows. 

\begin{figure}
\centerline{\includegraphics[width=6.5truecm]{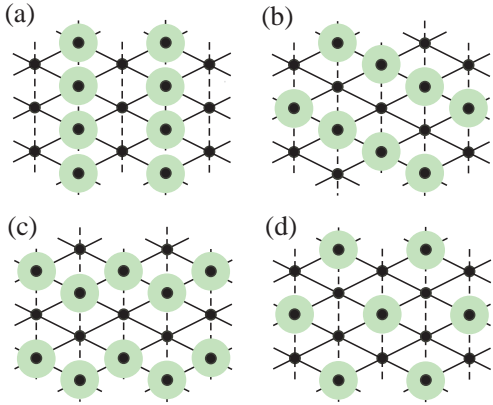}}
\vspace*{-1em}
\caption{(Color online) 
Charge ordered states 
in the quarter-filled extended Hubbard models 
on anisotropic triangular lattices: 
(a) vertical, (b) diagonal, and 
(c) horizontal stripe type states, 
and (d) the threefold state. 
Sites with/without colored circle represent the charge rich/poor sites. 
The solid and dotted bonds represent 
the ``$p$" ($p1 \sim p4$) and ``$c$" ($c1 \sim c3$) bonds in 
the $\theta$(or $\theta_{\rm d}$)-type and the $\alpha$-type structures in Fig.~\protect\ref{fig_polytypes}.  
}
\label{fig_stripes}
\vspace*{-1em}
\end{figure}

MF calculations on the anisotropic triangular lattice EHM 
were first performed by Seo~\cite{Seo00JPSJ} for different materials
with $\theta$ and $\alpha$-type structures. 
The $U$ and the $V_{ij}$ terms are treated within the standard MF treatment as  
$n_{i\sigma} n_{j\sigma'} \rightarrow n_{i\sigma} \langle n_{j\sigma'} \rangle 
+ \langle n_{i\sigma} \rangle n_{j\sigma'} 
-\langle n_{i\sigma} \rangle \langle n_{j\sigma'} \rangle$, 
by allowing large unit cell sizes to consider 
various CO states with several possible spin orders for each of them.  
In fact many self-consistent MF solutions are obtained 
and their ground state energies are compared. 
It is found that CO states with lowest energy 
show a variety as shown in Fig.~\ref{fig_stripes} depending on the parameters, 
which is in contrast to the square lattice case where the checkerboard pattern is stable. 
Mainly the ``stripe" type CO states are discussed in ref.~\ref{Seo00JPSJ}, 
shown in Fig.~\ref{fig_stripes}(a) $\sim$ (c), 
where the carrier densities on 
the charge rich and the charge poor sites 
are (in some cases approximately) $1/2 +\delta$ and $1/2 -\delta$, respectively.  
These can be considered as extensions of the Wigner crystal-type CO, 
which can lead to insulators in the strongly correlated regime. 
In general, the vertical stripe type pattern is favored 
by the Coulomb repulsion along the solid bonds in Fig.~\ref{fig_stripes}, 
while that along the dotted bonds stabilizes the diagonal and horizontal patterns. 

Let us start with the MF results for the $\theta$-type structure, eq.~(\ref{eqn:theta}), 
which has the highest symmetry among the polytypes. 
As noted in $\S$~\ref{subsec_COq2d}, 
nearly isotropic $V_c \simeq V_p$ is expected~\cite{TMori00BCSJ}, 
which gives rise to close competition between the different CO patterns. 
MF calculations~\cite{Seo00JPSJ} for eq.~(\ref{eqn:theta}) at $T=0$ 
considering the stripe type CO patterns as in Fig.~\ref{fig_stripes}(a) $\sim$ (c), 
have been performed for $t_p=-0.1$ eV, $t_c=0.01$ eV, 
taken from typical extended H\"{u}ckel parameters 
for $\theta$-ET$_2MM'$(SCN)$_4$ ($M$: Rb, Cs, Tl, etc., and $M'$: Zn, Co, etc.)~\cite{HMori98PRB},
and $U=0.7$ eV, while varying $V_p$ and $V_c$ near $0.3$ eV. 
The results show that either vertical or diagonal stripe type CO solution is stabilized 
for $V_c \lesssim V_p$ and $V_c \gtrsim V_p$, respectively, 
whereas the horizontal pattern has rather higher MF energies. 
In the actual $\theta$-type compounds which sustain this structure 
down to low temperatures, 
either vertical~\cite{Yakushi02PRB} or diagonal~\cite{Suzuki04PRB}
pattern of CO is indicated by optical measurements, 
consistent with the MF results. 
We note that the vertical stripe CO state here 
is equivalent to 
the checkerboard pattern in the $t_c=0$ and $V_c=0$ square lattice 
EHM discussed in $\S$~\ref{subsec_SqL}. 

In the region of $V_c \simeq V_p$, 
first in a study setting $t_{ij}=0$~\cite{TMori03JPSJ}  
and recently in a MF study~\cite{Kaneko06JPSJ}, 
a solution with a larger periodicity called as the ``threefold" state 
(Fig.~\ref{fig_stripes} (d)), 
which was not considered in ref.~\ref{Seo00JPSJ}, 
is found to have lower energy than the stripe-type CO states. 
In this state, the carrier densities on 
the charge rich and the charge poor sites 
are $1/2 +2\delta$ and $1/2 -\delta$, respectively 
(the $t_{ij}=0$ study~\cite{TMori03JPSJ} provides the extreme case of $\delta=1/4$, 
namely, carrier densities of 1 and 1/4). 
In this state, the carriers can avoid the occupancy in both 
neighboring pairs, $V_c $ and $ V_p$,  
therefore it is stabilized at $V_c \simeq V_p$, 
compared with the stripe-type CO states 
which cost loss in either one of these Coulomb repulsion terms. 
Such a long period state in fact appears as a short range order 
as found in X-ray scattering measurements~\cite{Watanabe99JPSJ,Watanabe04JPSJ} 
as diffusive rods (no interlayer coherence), 
in the conductive states of several members of $\theta$-ET$_2X$. 
However these are observed at different 2D wave vectors than in the studies above,  
which may be resolved by introducing 
longer range of Coulomb interactions in the EHM~\cite{KurokiPrivate}. 
In any case this threefold state is metallic since it is incompatible with the periodicity for quarter-filling, 
in contrast with the stripe-type CO states which is insulating in the strongly correlated regime, 
therefore these instabilites should be distinguished. 
\begin{figure}
\centerline{\includegraphics[width=7.5truecm]{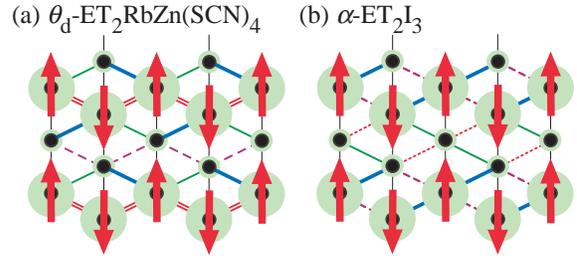}}
\vspace*{-1em}
\caption{(Color online) 
Schematic drawing of mean-field solutions~\cite{Seo00JPSJ}  
for the horizontal stripe-type charge ordered states 
in extended Hubbard model on anisotropic triangular lattices for 
(a) $\theta_{\rm d}$-ET$_2$RbZn(SCN)$_4$ and (b) $\alpha$-ET$_3$I$_3$. 
Different ``$p$" bonds are represented by different lines following Fig.~\ref{fig_polytypes}.  
} 
\vspace*{-1.5em}
\label{fig_stripesMF}
\end{figure}

Many $\theta$-type compounds exhibiting CO 
undergo a first order phase transition accompanying 
a structural transformation to a slightly dimerized structure, 
typically seen in several members of $\theta$-ET$_2MM'$(SCN)$_4$~\cite{HMori98PRB}: 
the $\theta_{\rm d}$-type structure, with a modulated molecular network as shown in Fig.~\ref{fig_polytypes}. 
MF calculations~\cite{Seo00JPSJ,Kaneko06JPSJ} under such structure with full anisotropy of $t_{ij}$ 
taken into account, but approximating $V_{ij}$ to be 
two kinds, $V_p$ and $V_c$, as in the calculation for the $\theta$-type structure, 
have been performed. 
The results suggest that a horizontal stripe-type solution 
as shown in Fig.~\ref{fig_stripesMF}(a) 
can have the lowest energy for $V_c \gtrsim V_p$. 
This is the pattern observed in these compounds,
such as in $\theta$-ET$_2$RbZn(SCN)$_4$ below the metal-insluator 
transition temperature $T_{\rm MI}$~\cite{Watanabe04JPSJ}. 
Such calculations have been performed by directly adopting the low temperature 
structure for simplicity, 
while the importance of the coupling to the lattice 
have been implied~\cite{Miyagawa00PRB,Seo00JPSJ,Clay02JPSJ}. 
It is interesting to start from the high temperature $\theta$-type structure 
and see whether the structural phase transition to $\theta_{\rm d}$-type
together with the CO transition could be reproduced, 
which awaits future studies.  

As for the $\alpha$-type structure adopting the values of $t_{ij}$ 
for $\alpha$-ET$_2$I$_3$, 
MF calculations~\cite{Seo00JPSJ} show that a horizontal stripe solution (Fig.~\ref{fig_stripesMF}(b)) 
is more stable than 
the other solutions in the $V_c \gtrsim V_p$ region, 
in contrast to the vertical stripe type solution found by Kino and Fukuyama~\cite{Kino95JPSJ} 
stabilized in the small $V_{ij}$ region. 
This horizontal stripe pattern is in fact in accordance with the experiments 
below the metal-insulator CO phase transition temperature in this material~\cite{KakiuchiPrivate}.  
It is noteworthy that, 
in both this salt and in the $\theta_{\rm d}$-type ET$_2X$ above, 
the MF calculations and the experiments agree with each other 
to a degree of the actual molecules having rich and poor carrier density 
in their rather complicated unit cells. 

Recently, Kobayashi {\it et al.}~\cite{Kobayashi05JPSJ,Katayama06JPSJ}
have extended such MF calculations on this compound, $\alpha$-ET$_2$I$_3$, 
to parameters under pressure,  
where a so-called narrow-gap semiconducting state is found by experiments~\cite{Tajima00JPSJ,TajimaThisVol}. 
The results show successive transitions from the CO insulating 
to the CO metallic phase, and then to a phase with a peculiar ``zero-gap" state,  
as a function of pressure.  
This zero-gap state is characterized by a dispersion of anisotropic Dirac fermions 
at a wave vector ${\boldsymbol  k^0}$, which crosses at the Fermi energy.
It is noteworthy that ${\boldsymbol  k^0}$ is an incommensurate wave vector 
due to an ``accidential degeneracy"~\cite{Herring37PR} 
naturally emerging in the band structure of this compound, 
but not related with the symmetry of the lattice structure. 
There are attempts~\cite{Katayama06JPSJ} to explain the anomalous transport properties
observed at high pressures~\cite{Tajima00JPSJ,TajimaThisVol}
based on such framework. 

The magnetic properties under the stripe-type CO states 
can be deduced from such MF results together with the picture mentioned above 
that we have learned from the 1D cases and the 2D square lattice case. 
The spin degree of freedom is expected to be 
described by (quasi-) 1D Heisenberg systems with spins on the charge rich sites along the stripes, 
then quantum fluctuation may destroy the antiferromagnetic 
spin order found in MF solutions (see Fig.~\ref{fig_stripesMF}). 
In fact, such behavior is typically observed in 
the measurements on $\theta$-ET$_2$RbZn(SCN)$_4$ and $\alpha$-ET$_2$I$_3$ at ambient pressure, 
both showing horizontal stripe type CO but absolutely different magnetic behavior. 
In the former, 
the CO is realized in the $\theta_d$-type structure, 
where the MF results show that the bonds between charge rich sites are all equivalent 
along one kind of bond, $p4$ in Fig.~\ref{fig_polytypes} 
(the double solid bonds in Fig.~\ref{fig_stripesMF}(a)), 
resulting in a uniform Heisenberg coupling $J_{p4}$ between spins along these stripes. 
On the other hand, in the latter compound, 
the horizontal stripe pattern results in an alternation 
of bonds along the charge rich sites, $p2$ and $p3$ in Fig.~\ref{fig_polytypes} 
(the thick solid and long dashed bonds in Fig.~\ref{fig_stripesMF}(b)), 
therefore spin singlet formation due to alternating $J_{p2}$ and $J_{p3}$ can be expected. 
These explain the difference seen in the magnetic susceptibility data: 
a Bonner-Fischer like low-dimensional localized spin behavior in the former~\cite{HMori98PRB} 
and a prominent spin-gap behavior in the latter~\cite{Rothaemel86PRB}. 
We note that in $\theta$-ET$_2$RbZn(SCN)$_4$, 
at lower temperature another phase transition takes place: 
a spin singlet formation due to the spin-Peierls mechanism  
along the stripes~\cite{HMori98PRB}. 
This is another experimental fact indicating   
the validity of describing the magnetic properties 
under stripe-type CO states as 1D Heisenberg models. 

\begin{figure}
\begin{center}
\centerline{\includegraphics[width=5.0truecm]{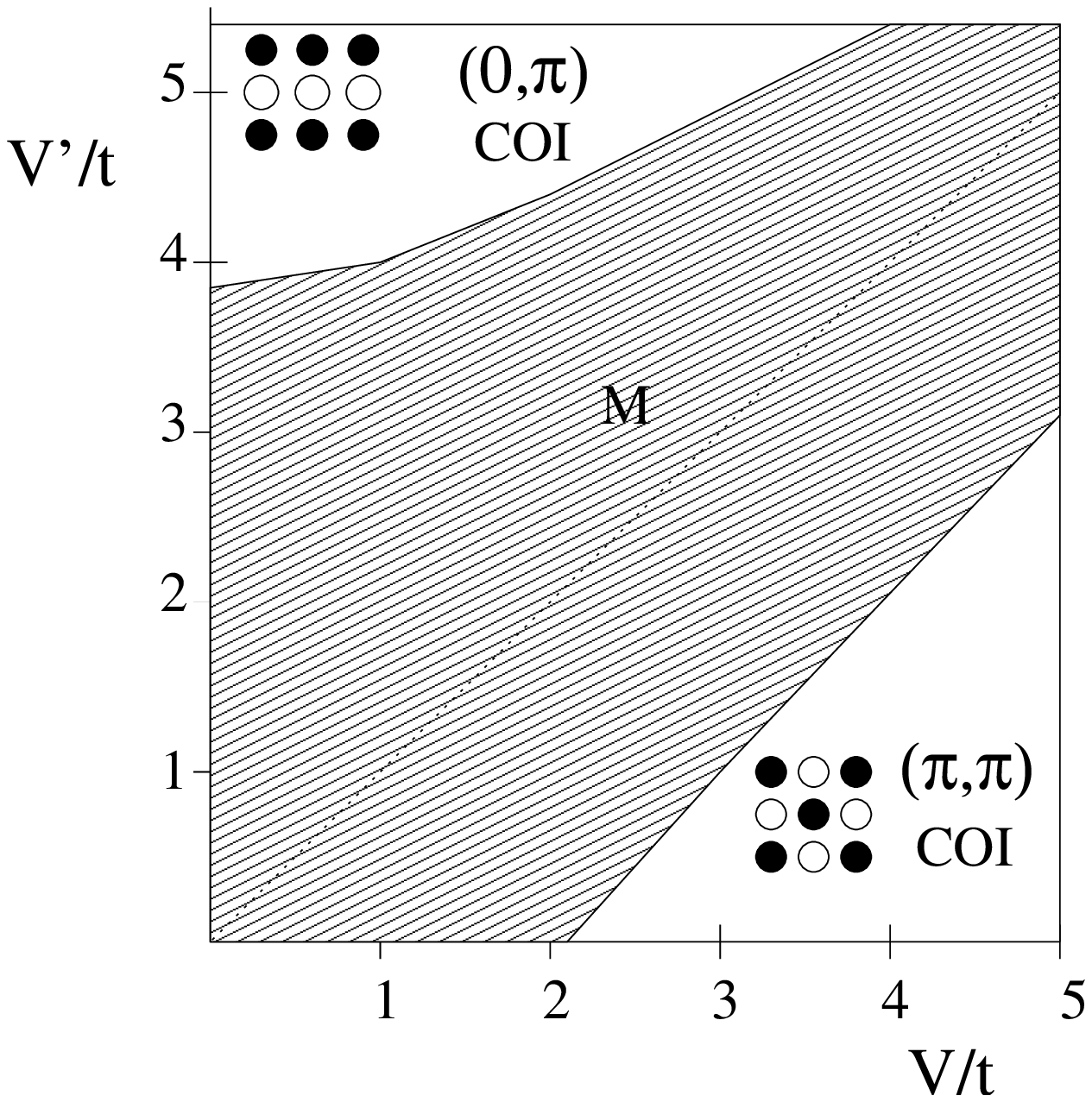}}
\vspace*{-2.5em}
\end{center}
\caption{Lanczos exact diagonalization ground state phase diagram 
of the quarter-filled extended Hubbard model on the $\theta$-type lattice structure for $U=10t$~\cite{Merino05PRB}. 
The figure is based on ``square lattice" representation: 
$V$ and $V'$ is respectively equavalent to $V_p$ and $V_c$ in eq.~(\ref{eqn:theta}), 
and $t_p=t$, $t_c=0$. 
The boundaries are 
extracted from the Drude weight calculation on a $L=16$ site cluster. 
The label M stands for metal, 
and COI for charge ordered insulator. 
[Reprinted figure with permission from 
J. Merino {\it et al}: Phys. Rev. B {\bf 71} (2005) 125111. 
Copyright (2005) by the American Physical Society.]
}
\vspace*{-1.5em}
\label{figphasedfrus}
\end{figure}

Although such procedure above is helpful in qualitatively understand the experiments, 
MF calculations cannot accurately describe situations 
where quantum fluctuations play crucial roles as mentioned in $\S$~\ref{sec_q1d}. 
This is indeed the case for the $\theta$-type structure near $V_c \simeq V_p$, 
as this is a situation under geometrical frustration introduced in $\S$~\ref{subsec_frustration}. 
In fact, computations including quantum fluctuation by Merino \et~\cite{Merino05PRB}
showed that the competition between different CO patterns can lead 
to a ``quantum melting" of CO, 
which is analogous to the case of the zigzag ladder model mentioned in $\S$~\ref{subsec_frustration}. 
The ground state phase diagram on the $V_c$-$V_p$ plane 
for fixed $U=10t$ is shown in Fig. \ref{figphasedfrus}, 
which is based on the Drude weight calculations on a $L=16$ site cluster
within the Lanczos ED technique explained in the previous subsection. 
There, a large metallic phase is stabilized 
in the region of $V_c \simeq V_p$, 
as in the 1D case we have seen in Fig.~\ref{fig_zigzag_phased}. 
The two different stripe-type CO states are stabilized 
in the less frustrated regions $V_c \gg V_p$ and $V_c \ll V_p$, 
which are, in contrast, consistent with the MF studies explained above~\cite{Seo00JPSJ}.  

Recently another study for this EHM on the $\theta$-type structure 
using the variational Monte Carlo method 
has been performed for much larger systems~\cite{WatanabeInprep}. 
In this method, 
variational states with CO patterns together with correlation effects are assumed 
and the variational energy is calculated numerically by Monte Carlo sampling~\cite{Yokoyama87JPSJ}. 
The features in the ED calculations above are reproduced, 
such as the robust metallic phase along the $V_p \simeq V_c$ line. 
In this metallic phase, the threefold periodic modulation of 
charge density is found as in the MF study~\cite{Kaneko06JPSJ} noted above, 
which, however, did not fit the $L=16$ cluster used in the ED study~\cite{Merino05PRB}. 
In the large $V_p \simeq V_c$ region, 
the threefold modulation is rather strong, 
which becomes weaker for small $V_p \simeq V_c$. 
There, different metallic states with stripe-type modulation, 
i.e., stripe-type CO metallic states, have very close variational energies. 
This indicates that there is large charge fluctuation~\cite{Merino05PRB,SeoInprep} 
including the threefold state due to the geometrical frustration. 

Combining the above results of different theoretical methods, 
one can consider that the horizontal stripe CO states found in the actual 
$\theta$-ET$_2X$ salts are realized by relaxing the frustration effect 
through the structural phase transition. 
In clear contrast, 
experiments show that members of $\theta$-ET$_2X$ which does not show such structural phase transition 
are rather conductive, but show a gradual increase of resistivity at low temperatures~\cite{HMori98PRB}. 
There are indications of a glassy CO state in $^{13}$C-NMR measurements~\cite{Kanoda05JP4}, 
which seems to be related with such large charge fluctuation. 
The additional disorder effect may be responsible for such behavior, 
which awaits to be understood. 

\subsection{Superconductivity}\label{subsec_SC}

\vspace{3pt}

Molecular conductors exhibit a variety of SC states 
as found in quasi-1D Bechgaard salts and quasi-2D ET salts~\cite{Ishiguro98Book}.
The issue of their mechanism, particularly 
the role of spin and charge fluctuations has attracted much attention~\cite{KurokiThisVol}. 
When an SDW or an antiferromagnetic phase is located 
next to the SC phase, 
it has been often asserted that the spin fluctuation is its origin. 
As an example, $\kappa$-ET$_2X$ shows SC next to an 
antiferromagnetic Mott insulating phase due to the dimerization~\cite{Miyagawa04CR}.  
Theoretical calculations for the half-filled Hubbard model on 
the anisotropic triangular lattice for the $\kappa$-type structure 
show that the SC state here has the $d_{x^2-y^2}$ symmetry mediated by antiferromagnetic 
spin fluctuation~\cite{Kino,Schmalian,Kondo}. 
Similarity to the high-$T_{\rm c}$ cuprates has been discussed~\cite{McKenzie97Science}, 
and in fact the $d$-wave state is 
stable when the lattice structure is continuously varied from the triangular lattice 
to the square lattice~\cite{OgatatJ,Kontani}. 

On the other hand, recent experiments on 2D $A_2B$ salts 
suggest the existence of a SC phase in the vicinity of the CO phase, 
such as in the unified phase diagram of 
$\theta$-type compounds~\cite{HMori98PRB} 
and SC is actually observed in $\theta$-(DIETS)$_2$Au(CN)$_4$ under uniaxial pressure~\cite{Tajima03JPSJ}. 
In $\alpha$-ET$_2$I$_3$ under uniaxial pressure, SC is even implied to show up 
in the presence of CO~\cite{Tajima02JPSJ}. 
These motivated theoretical studies 
to investigate possibilities of charge fluctuation as 
an origin of the SC state, 
based on the 2D EHM on the square lattice as well as on the anisotropic triangular lattice, 
which we will review in this subsection.

SC mediated by charge fluctuation was discussed 
first by Scalapino \et~\cite{Scalapino87PRB} 
for the 3D EHM on the cubic lattice. 
Relevance of this mechanism to the molecular conductors was pointed out 
first by Merino and McKenzie~\cite{Merino01PRL}, 
in the quarter-filled square lattice EHM 
by extending the slave-boson theory for the large-$N$ $t$-$V$ model~\cite{McKenzie01PRB} 
introduced in $\S$~\ref{subsec_SqL}. 
Later several authors~\cite{Kobayashi,Onari} 
have extended the weak coupling approach by Scalapino \et \ (see below) 
to the square lattice case as well. 
Their results all show that in the 2D case charge fluctuation due to the 
nearest-neighbor Coulomb repulsion $V$ 
can induce a singlet SC state with $d_{xy}$-wave symmetry. 

In the slave-boson theory for the large-$N$ $t$-$V$ model, 
the effective interaction $V_\textrm{eff}({\boldsymbol  q}={\boldsymbol k}-{\boldsymbol k'})$, 
acting between two quasiparticles with momentum ${\boldsymbol  k}$ and $-{\boldsymbol  k}$ 
which scatter to ${\boldsymbol  k'}$ and $-{\boldsymbol  k'}$,
can be calculated from the $1/N$ fluctuations 
around the MF solution at $N \rightarrow \infty$~\cite{Merino01PRL,Merino03PRB}.
As the ratio $V/t$ is increased, the potential
varies its shape developing singularities at $(\pm \pi/a, \pm \pi/a)$ 
toward the critical value $(V/t)_{\rm cr}$ for the occurance of checkerboard CO, 
at zero temperature (see $\S$~\ref{subsec_SqL}). 
This leads to an attraction in the channel with $d_{xy}$-symmetry, 
and the SC instability is estimated by the coupling averaged over the Fermi surface 
turning from positive to negative by decreasing temperature~\cite{Greco}. 
We note that the Fermi surface here has poor nesting, 
then the pairing mechanism is induced by the intersite Coulomb repulsion $V$. 

\begin{figure}
\centerline{\includegraphics[width=7truecm]{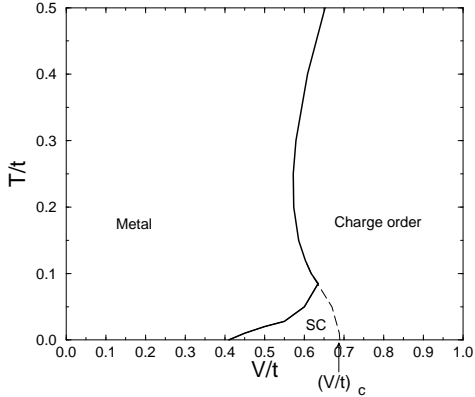}}
\vspace*{-1.5em}
\caption{Phase diagram of the large-$N$ $t$-$V$ model 
in the ($T/t$,$V/t$) plane 
showing competition between metallic, superconducting (SC), and charge
ordered phases~\cite{Merino01PRL}. The symmetry of the Cooper pairs in the
superconducting phase is d$_{xy}$, 
which is found near the quantum critical point $(V/t)_{\rm cr}$ 
separating the metallic and charge ordered phases. 
[Reprinted figure with permission from 
J. Merino and R. H. McKenzie: Phys. Rev. Lett. {\bf 87} (2001) 237002.
Copyright (2001) by the American Physical Society.]
}
\vspace*{-1.5em}
\label{figphased}
\end{figure}

The resulting $T$-$V$ phase diagram is shown in Fig. \ref{figphased}, 
where the $d_{xy}$-wave SC state is next to the CO phase discussed in $\S$~\ref{subsec_SqL}, 
near the critical point $(V/t)_{\rm cr}$. 
We note that a re-entrant behaviour for the CO phase transition temperature 
is found, 
which is also seen in the EHM in infinite dimension 
using the dynamical MF theory~\cite{Pietig99PRL}.
From the strong coupling viewpoint, the spin fluctuation associated
with the fourth order exchange process in eq.~(\ref{eqn:Jfourth}) 
couples the next-nearest neighbor sites antiferromagnetically (see Fig.~\ref{fig_CO2D}).
This spin interaction, in turn, cooperates with the charge fluctuation
in stabilizing the $d_{xy}$-wave SC state close to the CO transition~\cite{Greco}.

Search for a SC state near CO in the anisotropic triangular lattice EHM 
has been pursued by Tanaka \et~\cite{Tanaka}. 
They used the RPA treatment 
adopting $t_{ij}$ for $\theta$-(DIETS)$_2$[Au(CN)$_4$] as 
$t_c/t_p = 0.4$, 
and assumed an isotropic inter-site Coulomb repulsion $V_c=V_p=V$,  
in eq.~(\ref{eqn:theta}). 
In this approximation, the ladder-type diagrams containing $V$
are neglected, but a recent numerical calculation in the 
fluctuation exchange approximation retaining them~\cite{Onari} 
showed that the these terms do not change the results qualitatively.
The pairing interactions for the singlet and triplet channels are given by
\begin{align}
    \label{eq:s_pairing}
V^s({\boldsymbol  q}, \omega_l ) 
 &= U +V({\boldsymbol  q}) + \frac{3}{2}U^2 \chi_s ({\boldsymbol  q}, \omega_l )\nonumber
 \\
 &-(\frac{1}{2}U^2 +2U V({\boldsymbol  q})+2V({\boldsymbol  q})^2) \chi_c ({\boldsymbol  q}, \omega_l ),
\end{align}
\begin{align}
    \label{eq:t_pairing}
V^t({\boldsymbol  q}, \omega_l ) 
  &= V({\boldsymbol  q}) -\frac{1}{2}U^2 \chi_s ({\boldsymbol  q}, \omega_l )\nonumber \\
  &- (\frac{1}{2}U^2 +2U V({\boldsymbol  q})
                 +2V({\boldsymbol  q})^2) \chi_c ({\boldsymbol  q}, \omega_l ),
\end{align}
respectively, where 
$V({\boldsymbol  q})=2V(\cos q_x+\cos q_y + \cos (q_x+q_y))$ and $\omega_l$
is the Matsubara frequency ($x$ and $y$-directions are taken along the $p$ bonds in Fig.~\ref{fig_polytypes}). 
$\chi_s$ and $\chi_c$
are the spin and charge susceptibilities, respectively. 
They are calculated within RPA as,
$\chi_s ({\boldsymbol  q}, \omega_l ) = 
 \chi_0 ({\boldsymbol  q}, \omega_l )/(1-U\chi_0 ({\boldsymbol  q}, \omega_l )),
 \chi_c ({\boldsymbol  q}, \omega_l ) = 
 \chi_0 ({\boldsymbol  q}, \omega_l )
   /(1+(U+2V({\boldsymbol  q}))\chi_0 ({\boldsymbol  q}, \omega_l ))$,
where $\chi_0$ is the bare susceptibility.
We note that the terms proportional to $\chi_c$ in eqs. (\ref{eq:s_pairing})
and (\ref{eq:t_pairing}) represent effective pairing potentials due to
charge fluctuation. This charge fluctuation contributes equally to $V^s$ 
and to $V^t$ since it comes from the charge degrees of freedom.

\begin{figure}
\centerline{\includegraphics[width=6.5truecm]{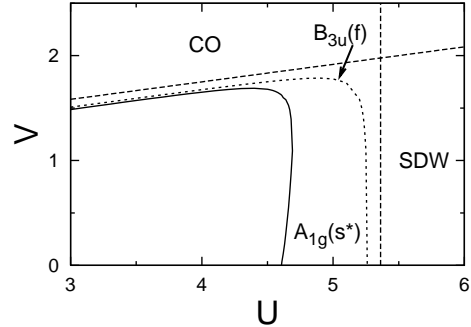}}
\vspace*{-0.5em}
\caption{ 
A weak coupling phase diagram for the quarter-filled extended Hubbard model for the $\theta$-type structure, 
on the $(U,V)$ plane at $T=0.01$ ($t_p$ is set as unity and $t_c=0.4$)~\cite{Tanaka}. 
The dashed lines correspond to the CO and SDW instabilities. 
The superconducting order parameter with
$B_{3u}$ symmetry is stabilized near the CO instability, although 
its eigenvalue is slightly smaller than that of the $A_{1g}$ symmetry.}
\label{fig_Tanaka4}
\end{figure}

To determine the onset of the SC state, 
the linearized $\acute{\rm E}$liashberg's equation in 
the weak coupling theory is solved. 
The obtained phase diagram on the $(U,V)$ plane is shown in 
Fig.~\ref{fig_Tanaka4} at a fixed temperature $T=0.01$.
$A_{1g}(s^*)$ represents the spin-singlet SC state 
with $A_{1g}$ symmetry, which is stabilized in the vicinity of 
both SDW (due to the nesting of Fermi surface) and CO (the threefold state 
discussed in $\S$~\ref{subsec_ATL}) instabilities. 
When $V\simeq 0$, the momentum dependence of the SC 
order parameter, $\Delta({\boldsymbol  k})$, becomes $d_{xy}$-like, 
similarly to the square lattice case~\cite{Merino01PRL,Kobayashi,Onari}. 

Near the CO instability, on the other hand, 
a spin-triplet SC with $B_{3u}$ symmetry is 
stabilized in addition to the spin-singlet state, although the 
eigenvalue is slightly smaller than that of the singlet pairing.
The momentum dependence of the order parameter $\Delta(\boldsymbol  k)$ 
shows that it is an $f$-wave pairing state since it changes 
the sign six times on the Fermi surface, 
which is similar to the isotropic triangular lattice case.~\cite{Tanaka2}
This SC state is stable because the momentum 
dependence of $V^t$ gives large attractive interactions at the 
wave vectors such as ${\boldsymbol  Q}=({2\pi}/{(3a)},{2\pi}/{(3a)})$.
Actually $\chi_c$ has a peak at ${\boldsymbol  Q}$ when $(U=3t_p,V=1.5t_p)$
leading to a peak in $V^t({\boldsymbol  q})$. 
We note that, similarly to the square lattice case, 
this momentum ${\boldsymbol  Q}$ is not due to the nesting
instability of the noninteracting Fermi surface:
$\chi_0({\boldsymbol  q})$ has no notable peak, and ${\boldsymbol  Q}$ is 
determined by the momentum dependence of $V({\boldsymbol  q})$. 
A recent variational Monte-Carlo calculation~\cite{WatanabeInprep}
have also found the $f$-wave SC state to be stabilized, 
which suggests that the SC phase is near but not next to the CO phase. 

The stabilities of the $d_{xy}$-wave pairing in the square lattice 
and the $f$-wave pairing in the anisotropic triangular lattice 
can also be understood from a naive real space picture. 
The nearest neighbor inter-site Coulomb repulsion repels electrons from the nearest-neighbor sites. 
The amplitude of the order parameter in real space becomes larger
at the four next-nearest-neighbor sites in the square lattice, 
while in the anisotropic triangular lattice there are six next-nearest-neighbor sites 
with large amplitude. 
These corresponds respectively to the $d_{xy}$ and the $f$-wave pairings. 

Let us briefly mention about SC found in $\alpha$-(ET)$_2$I$_3$ 
under uniaxial pressure which appears inside the CO phase; 
the SC phase transition takes place at a temperature below an upturn of the resistivity is observed, 
which is probably due to CO~\cite{Tajima02JPSJ}. 
RPA calculations have been performed by Kobayashi \et~\cite{Kobayashi05JPSJ}, 
based on the reconstructed Fermi surface 
under the the horizontal stripe-type CO in the metallic phase 
obtained in the MF results mentioned in \ref{subsec_ATL}. 
The results indicate that 
the SC phase can appear inside the middle CO metallic phase, 
in which small hole-pockets and electron-pockets both exist\cite{Kobayashi05JPSJ}. 
The suggested symmetry of this SC state is {\it s}-wave, 
of which the pairing interaction is mainly given by the spin fluctuation.  
This implies that in this case the charge fluctuation is suppressed 
due to the actual existence of CO, and the pairing instability 
arises from the newly formed Fermi surface.

\section{Related Topics}\label{sec_related}

\vspace{3pt}

In this section we briefly mention several issues related 
to CO in $A_2B$ systems that we have discussed. 
There are many transition metal oxides showing CO 
whose patterns can be identified as the Wigner-crystal type one. 
Well known examples are the perovskite type compounds~\cite{Imada98RMP}, 
such as 3D $AM$O$_3$ or quasi-2D $A_2M$O$_4$ 
with $M$ being a transition metal. 
When $A$ is half substituted by another ion with different valence, as $A_{1/2}A'_{1/2}$, 
$M$ can become mixed valent of half integer, 
namely, a ``quarter"-filled situation is realized. 
In such cases, an ``NaCl"-type CO for the 3D, 
or the checkerboard-type CO for the quasi-2D compounds is frequently observed. 
This CO state is actually behind the scene for 
the well-known collosal magnetoresistance effect (CMR) 
in perovskite manganites~\cite{Tokura00Science}. 
However, whereas the charge patterns reminds us of the CO states 
in the molecular systems, 
the driving force in these systems is usually rather involved. 
Orbital degeneracy in the $d$-electron under cubic/tetragonal crystal field, 
leading to the Jahn-Teller effect which is especially strong for the $e_g$ electron, 
frequently results in orbital ordering and makes the electronic state highly anisotropic. 
This together with strong electron correlation give rise to interesting, 
but at the same time complicated phase diagrams. 

Another transition metal oxide compound has been discussed to show a similar CO transition 
as in the molecular conductors: NaV$_2$O$_5$, 
which has been intensively studied both theoretically as well as 
experimentally~\cite{LemmensReview}. 
The transition first found in the magnetic susceptibility~\cite{Isobe96JPSJ}, 
from a behavior of paramagnetic 1D localized spin systems 
to a spin-gapped state at 35 K, was revealed to be due to the CO transtion~\cite{Ohama99PRB}. 
The average valence of the vanadium V$^{4.5+}$ produces a quarter-filled 
$d_{xy}$-band, which leads to an effective  2D quarter-filled model. 
It is the EHM, ${\cal H}_{\rm EHM}$ in eq.~(\ref{eqn:extHub}), on the so-called trestle lattice, 
where two-leg ladders along the $a$-axis are coupled in a zigzag way along the $b$-axis. 
A MF study~\cite{Seo98JPSJ} predicted a CO state with a ``zigzag" pattern, 
which is now confirmed experimentally~\cite{Ohwada05PRL}.  
As discussed in $\S$~\ref{sec_q2d}, such a 2D model is
difficult to study in a controlled way, 
and some authors studied the quarter-filled EHM on a two-leg ladder system~\cite{LemmensReview,Vojta01PRB}.  
In fact, ladders are also found to be realized in molecular systems~\cite{RoviraReview}, 
which is an interesting target for future studies. 

Recently, 
a triangular lattice system 
Na$_x$CoO$_2$ is attracting interest due to a SC state 
appearing when intercalated by water, H$_2$O~\cite{Takada03Nature}. 
At $x=0.5$, a stripe-type CO state is stabilized, 
which is discussed to be coupled to the ordering of Na$^+$ ions~\cite{Foo04PRL}. 
This reminds us of the CO state in $\theta$-ET$_2$RbZn(SCN)$_4$ 
where the CO transition couples to the lattice degree of freedom and 
relaxes the geometrical frustration by lowering 
the symmetry of the lattice. 
Similar theoretical approaches to those discussed in this review have been 
applied to the EHM~\cite{Watanabe05JPSJ} and the 
``$t$-$V$" model~\cite{Montrunich04PRB} appropriate for this compound. 

In the research field of molecular $A_2B$ materials, 
one important direction of evolution is the inclusion of 
$d$-electron spin introduced in the $B$ unit 
coupled to the organic $\pi$-electron on $A$, 
the so-called $\pi$-$d$ systems~\cite{KobayashiThisVol}. 
The role of CO here is not clear 
as it seems that most of such materials are realized in 
effective half-filled systems such as in the 
$\kappa$- and $\lambda$-type structures. 
It is desired that $\pi$-$d$ systems with a quarter-filled $\pi$-band 
would be investigated. 
It is to be noted that a theoretical work on 
slightly different molecular system 
TPP[FePcCN$_2$]$_2$ (Pc: pentacene), where 
the $S=1/2$ Fe$^{3+}$ ion is implanted in the donor $\pi$ molecule itself, 
showed that the interaction between localized spins 
would highly stabilize the CO state and  
predicted a novel ferromagnetic ground state~\cite{Hotta05PRL}. 

Many different kinds of molecular conductors are continuously and constantly synthesized. 
We expect that CO states will appear ubiquitously in the new materials as well. 
Sometimes the effects of CO may be secondary; one recent example is 
the case of $AB_x$ with $x$ being an incommensurate value 
close to 1/2, i.e., with an off quarter-filled band.  
An ``incommensurate Mott insulator"
has been theoretically proposed~\cite{Yoshioka05JPSJ}, 
possibly realized in recently synthesized (MDT-TS)(AuI$_2$)$_{0.441}$~\cite{Takimiya04ChemMater}, 
where the CO instability at quarter-filling indirectly 
controls the peculiar Mott transition of the system. 

\section{Summary}\label{sec_sum}

\vspace{3pt}

To summarize, we have reviewed 
theoretical studies on charge ordering and related phenomena 
in the 2:1 charge transfer molecular conductors expressed as $A_2B$. 
The charge ordered states are successfully described by 
extended Hubbard models 
which starts from the tight-binding model 
treating the anisotropy carefully enough 
and consider the electronic correlation, i.e., 
not only the on-site but also the inter-site Coulomb repulsion, 
and in some cases with the coupling to the lattice. 
The physics therein is rich, 
revealed by considerably numerous amount of studies, 
but still many issues arise from extensive experiments which anticipate future investigations. 
Encouraged by this success of understanding the apparently complex system 
based on constructing microscopic models and treating them by different theoretical techniques,  
frontiers of research in this field are just about to expand. 
For example, recent activities on the photo-induced phase transitions in 
$A_2B$ systems~\cite{Koshihara,TajimaThisVol} 
requires theoretical efforts with new methods to explore 
their nonequilibrium dynamics from the microscopic models. 
Furthermore, there is ambitious proposals~\cite{FukuyamaThisVol} to apply our knowledge to 
even more complex molecular assemblies such as bio-related materials. 

\section*{Acknowledgments} 

\vspace{3pt}

We would like to acknowledge all of our collaborators 
in the works on which this review is based on. 
Especially, we are grateful to 
H. Fukuyama, A. Greco,  R. H. McKenzie, Y. Suzumura, Y. Tanaka, and M. Tsuchiizu 
for continuous collaborations. 
We also acknowledge many experimentalists for informative and suggestive discussions. 
Obviously without these we would not be able to proceed such exciting years of studies.  


\end{document}